\documentclass[aps,prd,floats,showpacs,nofootinbib]{revtex4}
\usepackage{epsfig,url,graphicx}

\newcommand{\beq}{\begin{equation}}
\newcommand{\eeq}{\end{equation}}
\newcommand{\beqa}{\begin{eqnarray}}
\newcommand{\eeqa}{\end{eqnarray}}
\def\lsim{\lesssim}

\def\Msun{M_{\odot}{\ }}

\begin{document}

\title{Cosmological dark matter annihilations into $\gamma$-rays -- 
a closer look}

\author{Piero Ullio}
\affiliation{SISSA, via Beirut 4, I-34014 Trieste, Italy}
\author{Lars Bergstr\"om, Joakim Edsj\"o}
\affiliation{Department of Physics, Stockholm University,
AlbaNova, SE-106 91 Stockholm, Sweden}
\author{Cedric Lacey}
\affiliation{Department of Physics, Durham University
South Road, Durham DH1 3LE, England}

\begin{abstract}
We investigate the prospects of detecting weakly interacting massive 
particle (WIMP) dark matter by measuring the contribution to the
extragalactic gamma-ray radiation induced, in any dark matter halo and at
all redshifts, by WIMP pair annihilations into high-energy photons.
We perform a detailed analysis of the very distinctive spectral features 
of this signal, recently proposed in a short letter by three of the authors:
The gamma-ray flux which arises from the decay of $\pi^0$ mesons 
produced in the fragmentation of annihilation final states shows
a severe cutoff close to the value of the WIMP mass. An even more 
spectacular signature appears for the monochromatic gamma-ray components,
generated by WIMP annihilations into two-body final states containing 
a photon: the combined effect of cosmological redshift and absorption 
along the line of sight produces sharp bumps, peaked at the rest frame 
energy of the lines and asymmetrically smeared to lower energies. 
The level of the flux depends both on the particle physics scenario
for WIMP dark matter (we consider, as our template 
case, the lightest supersymmetric particle in a few supersymmetry 
breaking schemes), and on the question of how dark matter clusters. 
Uncertainties introduced by the latter are thoroughly discussed 
implementing a realistic model inspired by results of the state-of-the-art 
N-body simulations and semi-analytic modeling in the cold dark matter 
structure formation theory. We also address the question of the potential 
gamma-ray background originating from active galaxies, presenting a novel 
calculation and critically discussing the assumptions involved and the induced 
uncertainties. Furthermore, we apply a realistic model for the absorption 
of gamma-rays on the optical and near-IR intergalactic radiation field to 
derive predictions for both the signal and background. Comparing the two, 
we find that there are viable configurations, in the combined parameter
space defined by the particle physics setup and the structure formation
scenario, for which the WIMP induced extragalactic gamma-ray signal 
will be detectable in the new generation of gamma-ray telescopes such as 
GLAST.
\end{abstract}
\date{\today}
\maketitle
\section{Introduction}\label{sec:intro}

The accumulated evidence for the existence of large amounts of
nonbaryonic dark matter in the Universe is by now compelling (for
a review, see e.g.\ \cite{lbreview}). Data on the cosmic microwave
background (CMB)\cite{cmbr} and supernova observations \cite{supernovae} 
jointly fix the energy
fraction in the form matter and cosmological constant (or
something similar) to $\Omega_M\sim 0.3$ and $\Omega_{\Lambda}\sim
0.7$, respectively. At the same time, the CMB measurements limit
the contribution from ordinary baryons to less than $\Omega_B\sim
0.05$, which is in excellent agreement with big bang
nucleosynthesis. This means that non-baryonic matter has to make
up most of the matter in the Universe, $\Omega_{DM}\simeq
\Omega_M$. Incidentally, recent measurements of the large-scale
distribution of galaxies independently confirm $\Omega_M=0.27\pm
0.06$ \cite{licia}, giving further credence to these conclusions. 
The currently best estimate of $\Omega_M$ comes from a joint analysis 
of CMB and large scale structure data \cite{2dF} and gives 
$\Omega_Mh^2=0.115\pm0.009$ where $h$ is the Hubble constant in units 
of 100 km s$^{-1}$ Mpc$^{-1}$.

When it comes to the question of how the dark matter is
distributed on small, galactic and sub-galactic, scales the
situation is much less clear, however (for a review, see, e.g.,
\cite{moorerev}). After being subject to an extensive debate, 
with both theoretical and observational controversies, it seems that 
the Cold Dark Matter (CDM) model, with dark matter made of, e.g., 
weakly interacting massive particles (WIMPs), or the $\Lambda$CDM model, 
with a contribution from the cosmological constant, are in fair agreement 
with current observations, so that drastic modifications like strong 
self-interaction are not urgently called for (see, e.g., \cite{primack}).

Focusing on the CDM model with WIMPs as dark matter candidates, there is 
an obvious interest to use as much as possible of the knowledge of the 
distribution of CDM given through state-of-the-art N-body simulations. 
In particular, the distribution of dark matter plays a crucial role 
in most WIMP detection methods, and determines therefore the possibility of 
identifying the dark matter and pinning down its particle properties.

In this vein, we recently presented a short note \cite{beu} (hereafter BEU)
where, contrary to previous predictions~\cite{previous}, it was shown that 
in the hierarchical picture 
found in CDM simulations the cosmic $\gamma$-ray signal from WIMP 
annihilations may be at the level of current estimates of the extragalactic
$\gamma$-ray flux. 
In this paper we deal more carefully with the issues of the formation of 
structure in a CDM or, rather, $\Lambda$CDM universe,
investigating the sensitivity of the expected gamma-ray flux to different 
treatments of the structure formation process. We also address the question 
of the diffuse background flux expected from various types of active 
galaxies and compare its spectral features with those of the signal from 
WIMP annihilations. We consider several sample cases in a 
theoretically favored WIMP scenario, that of supersymmetric dark matter, 
and highlight the possibility to disentangle such signals from the 
background in future measurements of the the extragalactic $\gamma$-ray 
flux, in particular with the GLAST detector. Results for 
both the signal and background components are presented implementing 
a careful treatment of the absorption of high energy gamma-rays in the 
intergalactic space caused by pair production on the optical and infrared 
photon background.

The paper is organized as follows. In Sec.~\ref{sec:dm} we set up the 
general formalism for computing the dark matter induced gamma-ray flux.
In Sec.~\ref{sec:halo} we investigate the properties of dark matter
halos, on all scales of relevance to our problem, in various semi-analytical 
and numerical simulation scenarios. Implications for the WIMP induced
gamma-ray flux are discussed in Sec.~\ref{sec:flux}, while
in Sec.~\ref{sec:bkg}, we investigate the background problem, including
the effects of varying within present observational limits the slope of 
the energy spectrum of the gamma-ray emission from active galaxies.
We also check the effects of removing some more resolved 
point sources, as may be expected 
for the next generation of experiments.
In Sec.~\ref{sec:applications} we show a few examples of what signals can be 
expected for one of the favored WIMP candidates, the neutralino, and give 
an estimate of sensitivity curves for the GLAST detector. 
Sec.~\ref{sec:conclusions} contains our conclusions.

\section{The dark matter induced extragalactic gamma-ray flux}
\label{sec:dm}

There are several ways to compute the gamma-ray flux generated in
unresolved cosmological dark matter sources. In BEU the result was
derived by tracing the depletion of dark matter particles with
the Boltzmann equation. The approach we describe here, in which we
simply perform a sum of contributions along a given line of sight
(or better, a given geodesic), gives the same result but 
shows more directly the role 
played by structure in the Universe. We assume a standard homogeneous
and isotropic cosmology, described by the metric: \beq ds^2 = c^2
dt^2 - R^2(t) \left[ dr^2 + S_k^2(r) d\Omega^2 \right]\;, \eeq
where $d\Omega^2=d\theta^2+\sin^2\theta d\phi^2$ and where the
function $S_k(r)$ depends on the overall curvature of the
Universe: 
\beq 
  S_k(r)=\cases{r,& $k=0$;\cr \arcsin r,& $k=+1$;\cr
  {\rm arsinh} r,& $k=-1$.} 
\eeq 
In our applications, we will safely
use $k=0$ (i.e.\  we assume a flat geometry for the Universe). 
The angular interval $d\Omega=\sin\theta d\theta d\phi$
may, e.g., correspond to the angular acceptance of a given
detector. At redshift $z$, $d\Omega$ and the radial increment
$dr$ determine the proper volume: 
\beq 
  dV = \frac{[R_0 S_k(r)]^2\,R_0}{(1+z)^3} dr d\Omega\;. 
\eeq 
Let $d{\cal N}_{\gamma}/dE (E,M,z)$ be, on average, the differential energy
spectrum for the number of $\gamma$-rays emitted, per unit of
time, in a generic halo of mass $M$ located at redshift $z$.
Even for large $M$, this source can be safely regarded as
point-like and unresolved (with the upcoming generation of
gamma-ray telescopes, it might be possible to resolve the
dark-matter induced flux from galaxies in the local group, but
almost certainly not further out). Summing over all such sources
present in $dV$, we can find the number of photons emitted in this
volume and, say, in the time interval $dt$ and energy range
$(E,\,E+dE)$;
the emission process being isotropic, the corresponding number of photons
$dN_\gamma$ collected by a detector on earth with effective area $dA$ during
the time $dt_0$ and in the (redshifted) energy range $(E_0,\,E_0+dE_0)$,
is equal to:
\beq
  dN_\gamma = e^{-\tau(z,E_0)} \left[(1+z)^3 \, \int dM\;\frac{dn}{dM}(M,z)\,
  \frac{d{\cal N}_{\gamma}}{dE}\left(E,M,z\right)\right]\;
  \frac{dV\, dA}{4\pi [R_0 S_k(r)]^2} \, dE_0 \,dt_0\;,
\label{eq:infflux}
\eeq
where we applied the relation
$dt \,dE = (1+z)^{-1} dt_0 \, (1+z) dE_0 = dt_0 \,dE_0$, and we introduced
the halo mass function $dn/dM$, i.e.\ the comoving number density of bound
objects that have mass $M$ at redshift $z$ (the factor $(1+z)^3$ converts
from comoving to proper volume). The first factor  in Eq.~(\ref{eq:infflux})
is an attenuation factor which accounts for the absorption of $\gamma$-rays
as they propagate from the source to the detector: the main effect for GeV
to TeV $\gamma$-rays is absorption via pair production on the extragalactic
background light emitted by galaxies in the optical and infrared range.
Detailed studies of this effect, involving a modeling of galaxy and star
formation and a comparison with data on the extragalactic background
light, have been performed by several groups (see, e.g.,
\cite{abs1,abs2,SaSt98,deJaSt02,abs4,abs}). 
We take advantage of the results recently
presented by the Santa Cruz group~\cite{abs}; we implement an analytic 
parameterization of the optical depth $\tau$, as a function of both redshift 
and observed energy, which reproduces within about 10\% the values for this 
quantity plotted in Figs.~5 and 7 in Ref.~\cite{abs} ($\Lambda$CDM model 
labelled ``Kennicut''; the accuracy of the parameterization is much better 
than the spread in the predictions considering alternative 
models~\cite{abs}). For comparison, we have verified that the 
results presented in Salamon and Stecker~\cite{SaSt98} (their model in 
Fig.~6, with metallicity correction)
is in fair agreement with the model we are assuming as a reference model
in the energy range of interest in this work, i.e.\ below a few hundred GeV.

The estimate of the diffuse extragalactic $\gamma$-ray flux due to the
annihilation of dark matter particles is then obtained by summing over
all contributions in the form in Eq.~(\ref{eq:infflux}):
\beqa
  \frac{d\phi_{\gamma}}{dE_0} 
  \equiv \frac{dN_{\gamma}}{dA \,d\Omega \,dt_0 \,dE_0}
  & = & \frac{1}{4 \pi} \int dr \,R_0 e^{-\tau(z,E_0)}
  \int dM\;\frac{dn}{dM}(M,z)\,
  \frac{d{\cal N}_{\gamma}}{dE}\left(E_0\,(1+z),M,z\right) \nonumber \\
  & = & \frac{c}{4 \pi} \int dz \frac{e^{-\tau(z,E_0)}}{H_0\,h(z)}
  \int dM\;\frac{dn}{dM}(M,z)\,
  \frac{d{\cal N}_{\gamma}}{dE}\left(E_0\,(1+z),M,z\right)\;.
\label{eq:flux1}
\eeqa
where the integration along the line of sight has been replaced by one
over redshift, $H_0$ is the Hubble parameter, $c$ is the speed of light 
and $h$ depends on the cosmological model,
\begin{equation}
  h(z)=\sqrt{\Omega_M(1+z)^3+\Omega_K(1+z)^2+\Omega_\Lambda}.
\label{eq:hofz}
\end{equation}
In this work
we put the contribution from curvature $\Omega_K=0$, in agreement with
the prediction from inflation and with recent measurements
of the microwave background \cite{cmbr}. Taking
the limit in which all structure is erased and dark
matter is smoothly distributed at all redshifts, Eq.~(\ref{eq:flux1})
correctly reduces to the analogous formula derived with the Boltzmann
equation in BEU (Eq.~(4) therein).

\section{The properties of halos}\label{sec:halo} 

Three ingredients are needed to use Eq.~(\ref{eq:flux1}) for an actual
prediction of the $\gamma$-ray flux. We need to specify the WIMP pair 
annihilation cross section
and estimate
the number of photons emitted per annihilation, as well as 
the energy distribution of these photons: the choice of the particle
physics model fixes this element.
As photons are emitted in the annihilation of two WIMPs, the flux from
each source will scale with the square of the WIMP number density in the
source.
The second element needed is then the dark matter density profile
in a generic halo of mass $M$ at redshift $z$. Finally we need to know
the distribution of sources, i.e.\ we need an estimate of the halo mass
function.

Some insight on the latter two ingredients comes from the $\Lambda$CDM model
for structure formation: we outline here hypotheses and results entering
the prediction for the dark matter induced flux. We start with the mass
function for dark matter halos.

\subsection{The halo mass function}

Press-Schechter~\cite{PreSch} theory postulates that the
cosmological mass function of dark matter halos can be cast into
the universal form: 
\beq 
  \frac{dn}{dM} = \frac{\bar{\rho}_0}{M^2}
  \nu f(\nu) \frac{d\,log\nu}{d\,log M} 
  \label{eq:massfunc} 
\eeq
where $\bar{\rho}_0$ is the comoving dark matter background
density, $\bar{\rho}_0 \simeq \rho_c \Omega_M$ with $\rho_c$ being
the critical density at $z=0$. We introduced also the parameter 
$\nu \equiv \delta_{sc}(z)/ \sigma(M)$, defined as the ratio between 
the critical overdensity required for collapse in the spherical model 
$\delta_{sc}$ and the quantity $\sigma(M)$, which is the present, 
linear theory, rms density fluctuation in spheres containing a mean 
mass $M$. An expression
for $\delta_{sc}$ is given, e.g., in Ref.~\cite{ECF}. $\sigma(M)$
is related to the fluctuation power spectrum $P(k)$, see e.g.
Ref.~\cite{Peebles}, by: 
\beq 
  \sigma^2(M) \equiv \int d^3k \;
  \tilde{W}^2(k\,R) \, P(k) 
\eeq 
where $\tilde{W}$ is the top-hat
window function on the scale $R^3=3M/4\pi \bar{\rho}$ with
$\bar{\rho}$ the mean (proper) matter density. The power spectrum
is parametrized as $P(k) \propto k^n T^2(k)$; we fix the spectral
index $n=1$ and take the transfer function $T$ as given in the fit
by Bardeen et al.~\cite{Bardeenetal} for an adiabatic CDM model,
with the shape parameter modified to include baryonic matter according
to the prescription in, e.g.~\cite{Peacock}, Eq.\ (15.84) and (15.85).
Note that the fit we use agrees within 10\%
with the analytic result obtained for large $k$ in
Ref.~\cite{HUSU}, hence holds to the accuracy we are concerned
about for the small scales we will consider below. We normalize
$P$ and $\sigma$ by computing $\sigma$ in spheres of $R = 8/h$~Mpc
and setting the result equal to the parameter $\sigma_8$ ($h$ is the usual 
Hubble constant in units of 100 km s$^{-1}$ Mpc$^{-1}$).

In Eq.~(\ref{eq:massfunc}) $f(\nu)$ is known as the multiplicity function;
we implement the form found in the ellipsoidal collapse model~\cite{SMT}:
\beq
  \nu f(\nu) = 2 A \left(1+\frac{1}{\nu'^{2q}}\right)
  \left(\frac{\nu'^{2}}{2\pi}\right)^{1/2}
  \exp\left(-\frac{\nu'^{2}}{2}\right)
  \label{eq:nuofnu}
\eeq
where $\nu'=\sqrt{a}\nu$, and the parameters $q = 0.3$ and $a=0.707$ are
derived by fitting Eq.~(\ref{eq:massfunc}) to the N-body simulation results
of the Virgo consortium~\cite{virgo}, while $A$ is fixed
by the requirement that all mass lies in a given halo,
i.e.\ $\int d\nu f(\nu) = 1$ or $\int dM \;M dn/dM = \bar{\rho}_0$.
Eq.~(\ref{eq:nuofnu}) reduces to the form originally proposed in
Press-Schechter theory and valid for spherical collapse if
$a=1$, $q = 0$ and $A=0.5$.

\begin{figure}[t]
\centerline{\includegraphics[width=0.49\textwidth]{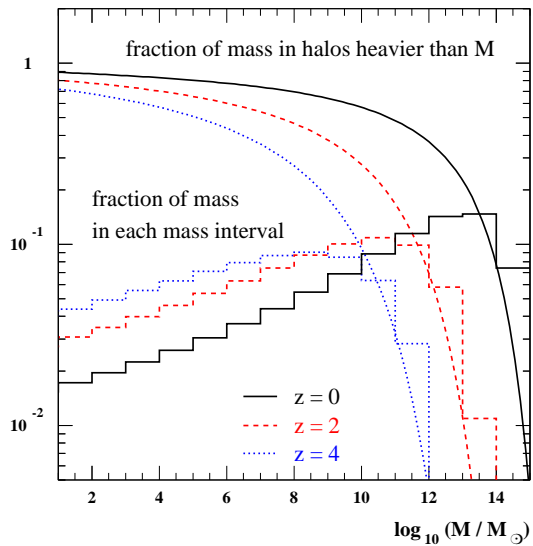}}
\caption{Fraction of total mass provided by objects heavier than a 
given mass $M$ (upper curves) or within 14 decades in mass 
(lower histograms) at three different redshifts and for the mass 
function as derived in the ellipsoidal collapse model.}
\label{fig:fig1}
\end{figure}

To give the reader a feeling for what the distribution of mass is, as
predicted by the halo mass function we are considering here, in
Fig.~\ref{fig:fig1} we plot the fraction of the total mass in halos
heavier than $M$, and the fraction per mass decade, for three different
redshifts, $z$ equal to 0, 2 and 4, and for our default choice of
cosmological model: $\Omega_M = 0.3$,
$\Omega_{\Lambda} = 0.7$, $h = 0.7$, $\Omega_b = 0.022/h^2$ and
$\sigma_8 = 0.73$~\cite{2df}. Note the peak in the distribution at
$M \sim 10^{12}-10^{13}\, \Msun$ for $z=0$ rapidly moving to lower
masses for larger redshifts; note also that the low mass tails are
not very steep, with only 89\% (81\%) of the total mass in structures
heavier than $10\, \Msun$ at $z=0$ ($z=2$). These numbers get slightly
larger if one applies the spherical collapse model instead of the
ellipsoidal model we have considered here.

\subsection{The density profile in dark halos}

In the $\Lambda$CDM model for structure formation, dark matter
halos are assumed to form hierarchically bottom-up via
gravitational amplification of initial density fluctuations. 
Small structures merge into larger and larger halos and final 
configurations are self-similar, with a smooth dark matter 
component and, possibly, a small fraction of the total mass
in subhalos which have survived tidal stripping.
We neglect for the moment eventual substructure, whose role
on the WIMP induced signal is discussed in the next Section. 
N-body simulations seem to indicate that dark matter density profiles can be
described in the form 
\beq 
  \rho(r)= \rho^{\prime} \,g(r/a)\;,
  \label{eq:haloprof} 
\eeq 
where $a$ is a length scale and
$\rho^{\prime}$ the corresponding density. The function $g(x)$ is
found to be more or less universal over the whole mass range of the simulated
halos, although different functional forms have been claimed in
different simulations: we will consider the result originally
proposed by Navarro, Frenk and White~\cite{NFW} (hereafter NFW
profile), 
\beq
g_{\rm NFW}\left(x\right)={1\over x \left(1+x\right)^2},
\eeq
 supported also by
more recent simulations performed by the same group~\cite{nfwnew},
and the result found in the higher resolution simulation (but with
fewer simulated halos) by Moore et al.~\cite{moore} (hereafter
Moore profile), 
\beq
g_{\rm Moore}\left(x\right)={1\over x^{1.5}\left(1+x^{1.5}\right)}. 
\eeq
The
two functional forms have the same behavior at large radii and
they are both singular towards the center of the halo, but the
Moore profile increases much faster than the NFW profile
(non-universal forms, with central cusp slopes depending on
evolution details have been claimed as well~\cite{klypin}).
There have been a number of reports in the literature arguing that
the rotation curves of many small-size disk galaxies rule out
divergent dark matter profiles, see, e.g.,~\cite{FP,moorerot}
(note however that this issue is not settled yet, see,
e.g.,~\cite{VDB}), while they can be fitted by profiles with a
flat density core. We consider then here as a third alternative
functional form the Burkert profile~\cite{Burkert}, 
\beq
g_{\rm B}\left(x\right)={1\over \left(1+x\right)\left(1+x^2\right)},
\eeq
 which has been shown to be adequate
to reproduce a large catalogue of rotation curves of spiral
galaxies~\cite{BS}.

Rather than by $a$ and $\rho^{\prime}$, it is useful to label a dark
matter profile by its virial mass $M$ and concentration parameter $c_{vir}$.
For the latter, we adopt here the definition by Bullock et al.~\cite{Bullock}:
let the virial radius $R_{vir}$ of a halo of mass $M$ at redshift $z$ be
defined as the radius within which the mean density of the halo is
$\Delta_{vir}$ times the mean background density $\bar{\rho}(z)$ at that
redshift: 
\beq
M \equiv  {4\pi\over 3} \Delta_{vir} \bar{\rho}(z)\, R_{vir}^3.
\eeq
We take the virial overdensity to be approximated by the expression~\cite{BN},
valid in a flat cosmology,
\beq
\Delta_{vir} \simeq {(18\pi^2 + 82x
- 39 x^2)\over \Omega_M(z)}
\eeq
with $x \equiv \Omega_M(z) -1$, ($\Delta_{vir} \simeq 337$
for $\Omega_M=0.3$ at $z=0$).
The concentration parameter is then defined as
\beq
  c_{vir} = \frac{R_{vir}}{r_{-2}}
\eeq
with $r_{-2}$ the radius at which the effective logarithmic slope of
the profile is $-2$, i.e.\ it is the radius set by the equation
$d/dr\left.\left(r^2 g(r)\right)\right|_{r=r_{-2}} = 0$.
This means that $r_{-2} = a$ for the NFW profile, while $x_{-2} \equiv r_{-2}/a$
is equal to about 0.63 for the Moore profile and to 1.52 for the
Burkert profile. Note that these definitions of $R_{vir}$ and $c_{vir}$
differ from those adopted in Ref.~\cite{NFW} and Ref.~\cite{ENS}.

\begin{figure}[t]
\centerline{\includegraphics[width=0.49\textwidth]{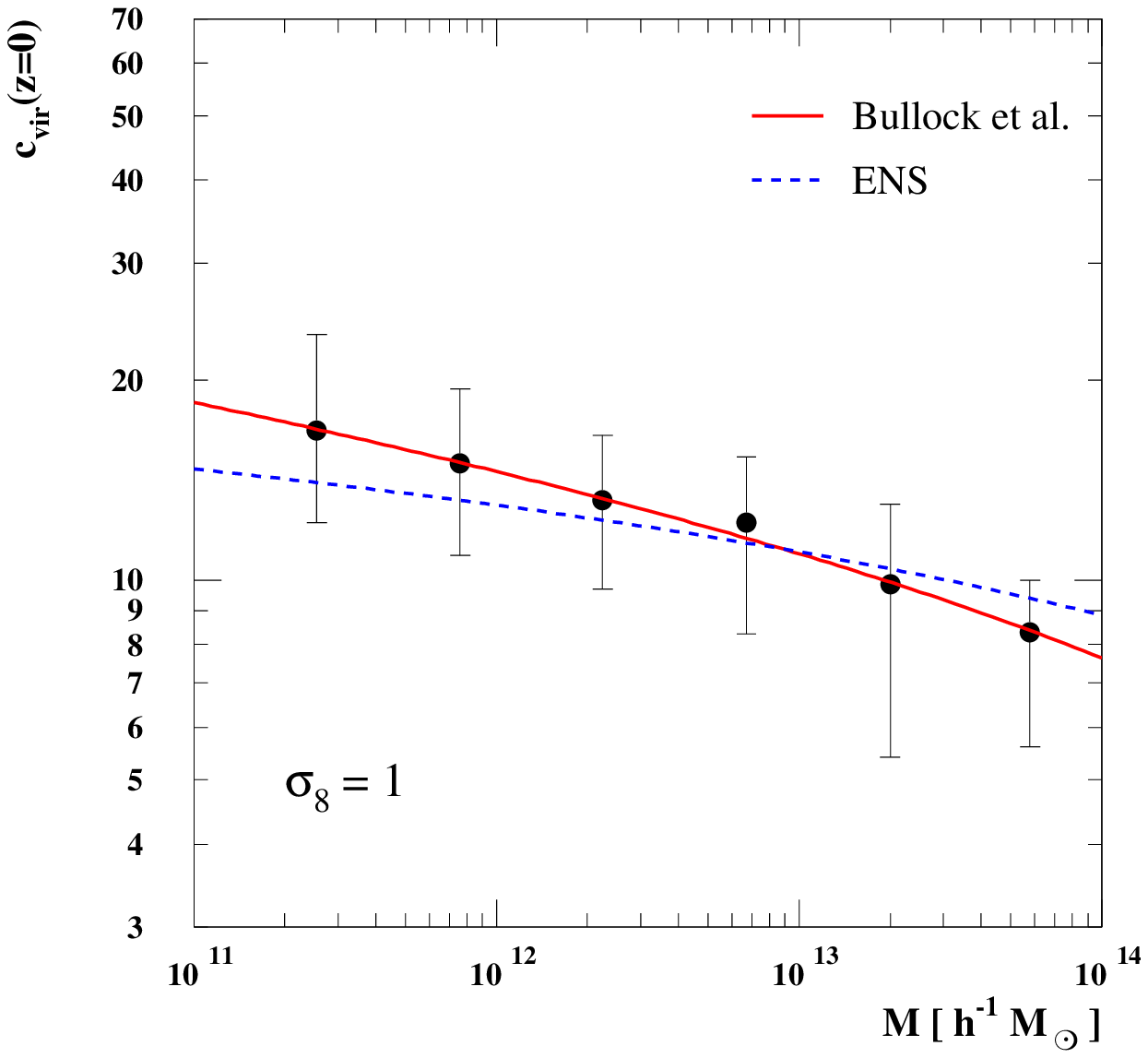}
\includegraphics[width=0.49\textwidth]{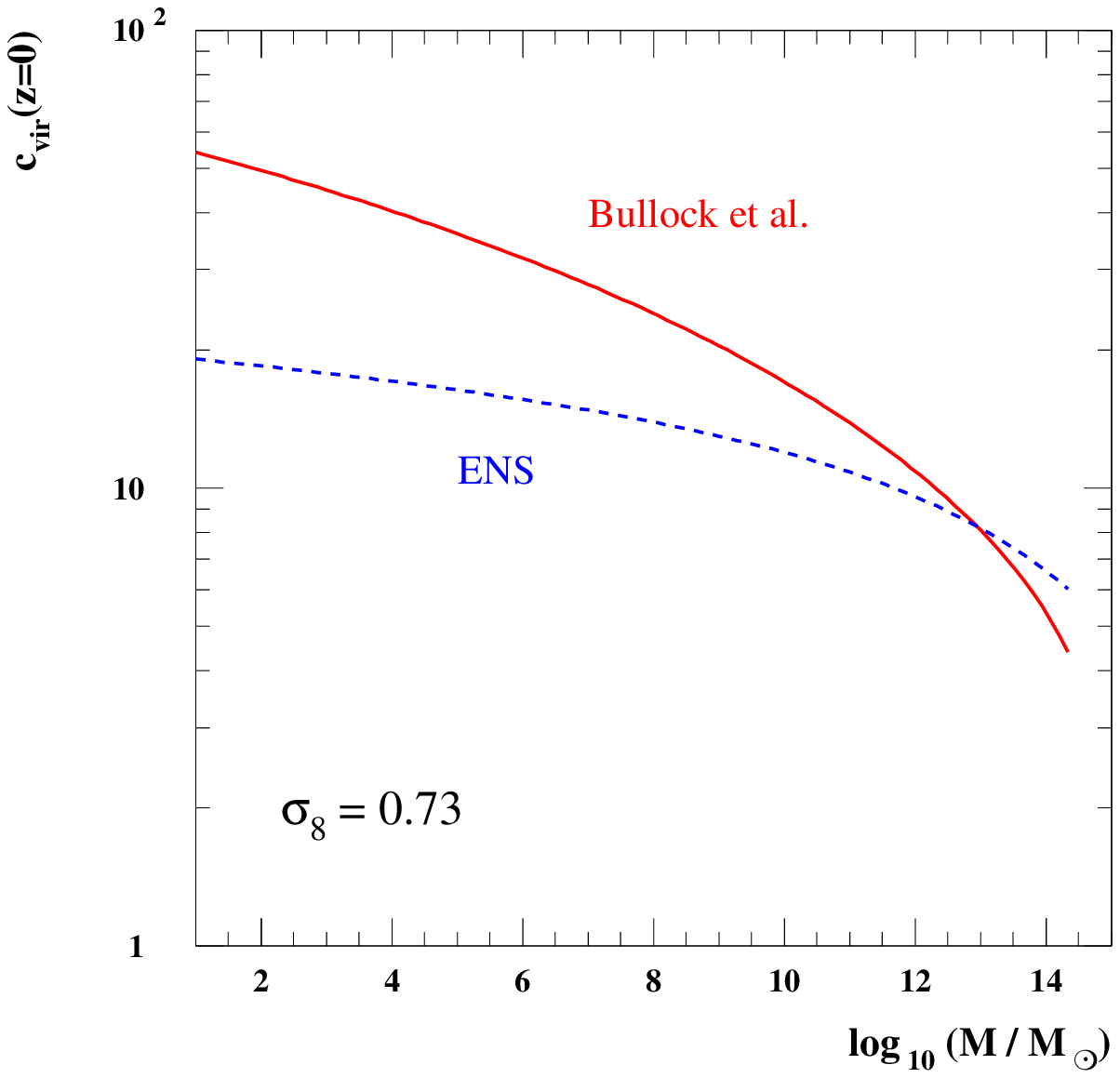}}
\caption{Concentration parameter versus mass for halos of mass $M$ at $z=0$.
On the left-hand panel we reproduce from Ref.~\protect{\cite{Bullock}}
the behavior found in a large sample of simulated halos, with a binning in 
mass in which each marker represents the peak in the distribution and 
the relative bar its 68\% width; the trend is reproduced with the toy 
models proposed in Ref.~\protect{\cite{Bullock}} itself (Bullock et al.) 
and in Ref.~\protect{\cite{ENS}} (ENS). On the right-hand side, 
we show an extrapolation of $c_{vir}$ to the whole mass range we need 
to include in our analysis according to the two toy models.}
\label{fig:fig2}
\end{figure}

After identifying the behavior in Eq.~(\ref{eq:haloprof}), Navarro
et al.\ noticed also that, for a given  cosmology, the halos in
their simulation at a given redshift show a strong correlation
between $c_{vir}$ and $M$~\cite{NFW}, with larger concentrations
in lighter halos. This trend may be intuitively explained by the
fact that low-mass halos typically collapsed earlier, when the
Universe was denser. Bullock et al.~\cite{Bullock} confirmed this
behavior with a larger sample of simulated halos and propose a
toy model to describe it, which improves on the toy model
originally outlined in~\cite{NFW}: On average, a collapse redshift $z_c$ is
assigned to each halo of mass $M$ at the epoch $z$ through the
relation $M_{\star}(z_c) \equiv F M$, where the typical collapsing
mass $M_{\star}$ is defined implicitly by
$\sigma\left(M_{\star}(z)\right)= \delta_{sc}(z)$ and is
postulated to be a fixed fraction $F$ of $M$ (following
Ref.~\cite{Wechsleretal} we choose $F=0.015$). The density of the
Universe at $z_c$ is then associated with a characteristic density
of the halo at $z$; it follows that, on average, the concentration 
parameter is given by:
\beq 
  c_{vir}(M,z) = K \frac{1+z_c}{1+z}  =
  \frac{c_{vir}(M,z=0)}{(1+z)} 
\label{cvir} 
\eeq 
where $K$ is a
constant (i.e.\ independent of $M$ and cosmology) to be fitted to
the results of the simulations. Bullock et al.~\cite{Bullock} show
that this toy model reproduces rather accurately the dependence of $c_{vir}$
found in the simulations on both $M$ and $z$. We
reproduce this fit at $z=0$ in Fig.~\ref{fig:fig2} (left panel,
solid line); ``data'' points and relative error bars are taken
from~\cite{Bullock} and just represent a binning in mass of
results in their simulated halos: in each mass bin, the marker and the
error bars correspond, respectively, to the peak and the 68\% width
in the $c_{vir}$ distribution. We determine $K$ with a best fitting
procedure in the cosmology $\Omega_M = 0.3$, $\Omega_{\Lambda} = 0.7$, 
$h = 0.7$ and $\sigma_8 = 1$ adopted in the N-body simulation referred to, 
and then use this value to estimate the mean $c_{vir}$ in other cosmologies;
we find $K = 4.4$. 
Finally, following again Bullock et al.~\cite{Bullock},
we assume that, for a given $M$, the distribution of concentration 
parameters ${\cal{P}}$ is log-normal with a $1 \sigma$ deviation 
$\Delta(\log_{10} c_{vir})$ around the mean, independent of $M$ and cosmology;
we take $\Delta(\log_{10} c_{vir})= 0.2$.

An alternative toy-model to describe the relation between
$c_{vir}$ and $M$ has been discussed by Eke, Navarro and
Steinmetz~\cite{ENS} (hereafter ENS model): The relation they
propose has a similar scaling in $z$, with however a different
definition of the collapse redshift $z_c$ and a milder dependence
of $c_{vir}$ on $M$. In our notation, they define of $z_c$ through
the equation: 
\beq 
  D(z_c) \sigma_{\rm eff}(M_p)= 
  {1 \over C_{\sigma}} 
\label{zcens} 
\eeq 
where $D(z)$ represents the linear
theory growth factor, and $\sigma_{\rm eff}$ is an `effective'
amplitude of the power spectrum on scale $M$: 
\beq 
  \sigma_{\rm eff}(M)=\sigma(M) \, 
  \left(-\frac{d\ln(\sigma)}{d \ln(M)}(M)\right) 
  = - \frac{d\sigma}{dM} M 
\label{sigeff} 
\eeq
which modulates $\sigma(M)$ and makes $z_c$ dependent on both the
amplitude and the shape of the power spectrum, rather than just on
the amplitude as in the toy model of Bullock et al. Finally, in
Eq.~(\ref{zcens}), $M_p$ is assumed to be the mass of the halo
contained within the radius at which the circular velocity reaches
its maximum, while $C_{\sigma}$ is the parameter (independent on
$M$ and cosmology) which has to be fitted to the simulations. With
this definition of $z_c$ it follows that, on average, $c_{vir}$ can
be be expressed as: 
\beq c_{vir}(M,z) =
  \left(\frac{\Delta_{vir}(z_c)\,\Omega_M(z)}
  {\Delta_{vir}(z)\,\Omega_M(z_c)}\right)^{1/3} 
  \frac{1+z_c}{1+z}\;.
\label{cvirens} 
\eeq 
As we already mentioned the dependence of
$c_{vir}$ on $M$ as given in the equation above is weaker than in
the Bullock et al.\ toy-model. Our best fitting procedure
gives $C_{\sigma} = 76$ and the behavior in
Fig.~\ref{fig:fig2} (left panel, dashed line), which reproduces
the N-body ``data'' fairly well, 
with values not very far from those obtained in the Bullock et al.\ model 
within the range of simulated masses, and possibly just a slight 
underestimate of the mean value in the lighter mass end.

On the other hand, the extrapolation outside the simulated mass
range can give much larger discrepancies as shown in the right
panel of Fig.~\ref{fig:fig2}. Solid lines are for the same models
as those shown in the left panel ($K$ and $C_{\sigma}$ from the
data fit in the left panel), with just $\sigma_8$ set equal to our
preferred value, $\sigma_8 = 0.73$. When going to small $M$,
$c_{vir}$ increases in both cases, but the growth in the model of 
Bullock et al. is much faster than in the ENS model;
we will show explicitly how this uncertainty propagates to the prediction
of the dark matter induced $\gamma$-ray flux. The sensitivity of
our results to the choice of cosmological parameters is generally 
much weaker: The largest effect is given by the overall linear scaling 
of $c_{vir}(M,z)$ with $\sigma_8$. There is also the possibility to
change the cosmological model by including other dark components;
we are not going to discuss any such case in detail, we just mention 
that a neutrino component at the level of current upper limits is not 
going to change severely our picture, while a substantial warm dark 
matter component may play a crucial role if $z_c$ is indeed defined 
according to the ENS prescription.

\section{WIMP induced flux: role of structures and spectral features} 
\label{sec:flux}

We are now ready to write explicitly the term $d{\cal N}_{\gamma}/dE$ 
introduced above and to derive the formula for the
flux. The differential energy spectrum for the number of photons
emitted inside a halo with mass $M$ at redshift $z$ is 
\beqa
  \frac{d{\cal N}_{\gamma}}{dE} (E,M,z) & = & \frac{\sigma v}{2}
  \frac{dN_{\gamma}(E)}{dE} \int dc^{\,\prime}_{vir}\; 
  {\cal{P}}(c^{\,\prime}_{vir})
  \left(\frac{\rho^{\prime}}{M_{\chi}}\right)^2 
  \int d^3r\; g^2(r/a) \nonumber \\
  & = &\frac{\sigma v}{2} \frac{dN_{\gamma}(E)}{dE}
  \frac{M}{M_{\chi}^2} \, \frac{\Delta_{vir}\bar{\rho}}{3}\,
  \int dc^{\,\prime}_{vir}\; {\cal{P}}(c^{\,\prime}_{vir})
  \frac{(c^{\,\prime}_{vir}\,x_{-2})^3}
  {\left[I_1(c^{\,\prime}_{vir}\,x_{-2})\right]^2} \,
  I_2(x_{min},c^{\,\prime}_{vir}\,x_{-2})
\label{eq:dnde}
\eeqa
where $\sigma v$ is the WIMP annihilation rate, $dN_{\gamma}(E)/dE$
is the differential gamma-ray yield per annihilation and $M_{\chi}$ is
the WIMP mass. Note that in previous literature, the prefactor $\sigma v/2$ 
has often been erroneously taken as $\sigma v$. The derivation 
based on the Boltzmann equation in BEU adds to our confidence that the factor 
should be $\sigma v/2$. In Eq.~(\ref{eq:dnde}) we applied the 
definition of $R_{vir}$ and introduced the integrals
\beq
  I_n(x_{min},x_{max}) = \int_{x_{min}}^{x_{max}} dx\, x^2 g^n(x)\;.
\eeq
with the lower limit of integration $x_{min} = r_{min}/a$ set, 
in a singular halo profile, by WIMP self annihilations, i.e.\ roughly
by $\rho(r_{min}) \simeq m_{\chi}/[\sigma v\; (t_0 - t_c)]$, where
$t_0$ is the age of the Universe and $t_c$ is the collapse time for the 
halo under investigation. To include all sources labeled by their mass $M$,
we averaged over the log-normal distribution ${\cal{P}}(c^{\,\prime}_{vir})$ 
centered on $c_{vir}$ as given in Eq.~(\ref{cvir}) or~(\ref{cvirens}).

Inserting Eq.~(\ref{eq:dnde}) into Eq.~(\ref{eq:flux1}), we find that the 
gamma-ray flux is
\beq
  \frac{d\phi_{\gamma}}{dE_0} = \frac{\sigma v}{8 \pi} \frac{c}{H_0}
  \frac{\bar{\rho}_0^2}{M_{\chi}^2} \int dz\;(1+z)^3 \frac{\Delta^2(z)}{h(z)}
  \frac{dN_{\gamma}(E_0\,(1+z))}{dE} e^{-\tau(z,E_0)} \;,
\label{eq:flux2}
\eeq
where we have defined
\beq
  \Delta^2(z) \equiv \int dM \frac{\nu(z,M) f\left(\nu(z,M)\right)}{\sigma(M)}
  \left|\frac{d\sigma}{dM}\right| \Delta_M^2(z,M)
\label{eq:D2}
\eeq
and the quantity
\beq
  \Delta_M^2(z,M) \equiv
  \frac{\Delta_{vir}(z)}{3}\,\int dc^{\,\prime}_{vir}\; 
  {\cal{P}}(c^{\,\prime}_{vir})
  \frac{I_2(x_{min},c^{\,\prime}_{vir}(z,M)\,x_{-2})}
  {\left[I_1(x_{min},c^{\,\prime}_{vir}(z,M)\,x_{-2})\right]^2} 
  (c^{\,\prime}_{vir}(z,M)\,x_{-2})^3 \;.
\label{eq:D2M}
\eeq
(Note that this definition differs from that in BEU \cite{beu} by a 
factor $1/(1+z)^3$. The advantage of the present definition is that 
$\Delta_M^2(z,M)=1$ if all matter is at the mean density for redshift
$z$.)
In early estimates of the WIMP induced extragalactic $\gamma$-ray 
flux, see, e.g.,~\cite{previous}, the role of structure was not appreciated
and the dark matter distribution was assumed to be described simply
by the mean cosmological matter density $\rho(z) = \rho_c \Omega_M (1+z)^3$.
Compared to this picture, $\Delta_M^2(z,M)$ gives the average enhancement
in the flux due to a halo of mass $M$, while $\Delta^2(z)$ is the sum over
all such contributions weighted over the mass function. As we will see, 
the enhancement of the annihilation rate due to structure amounts to several 
orders of magnitude.

\begin{figure}[t]
\centerline{\includegraphics[width=0.49\textwidth]{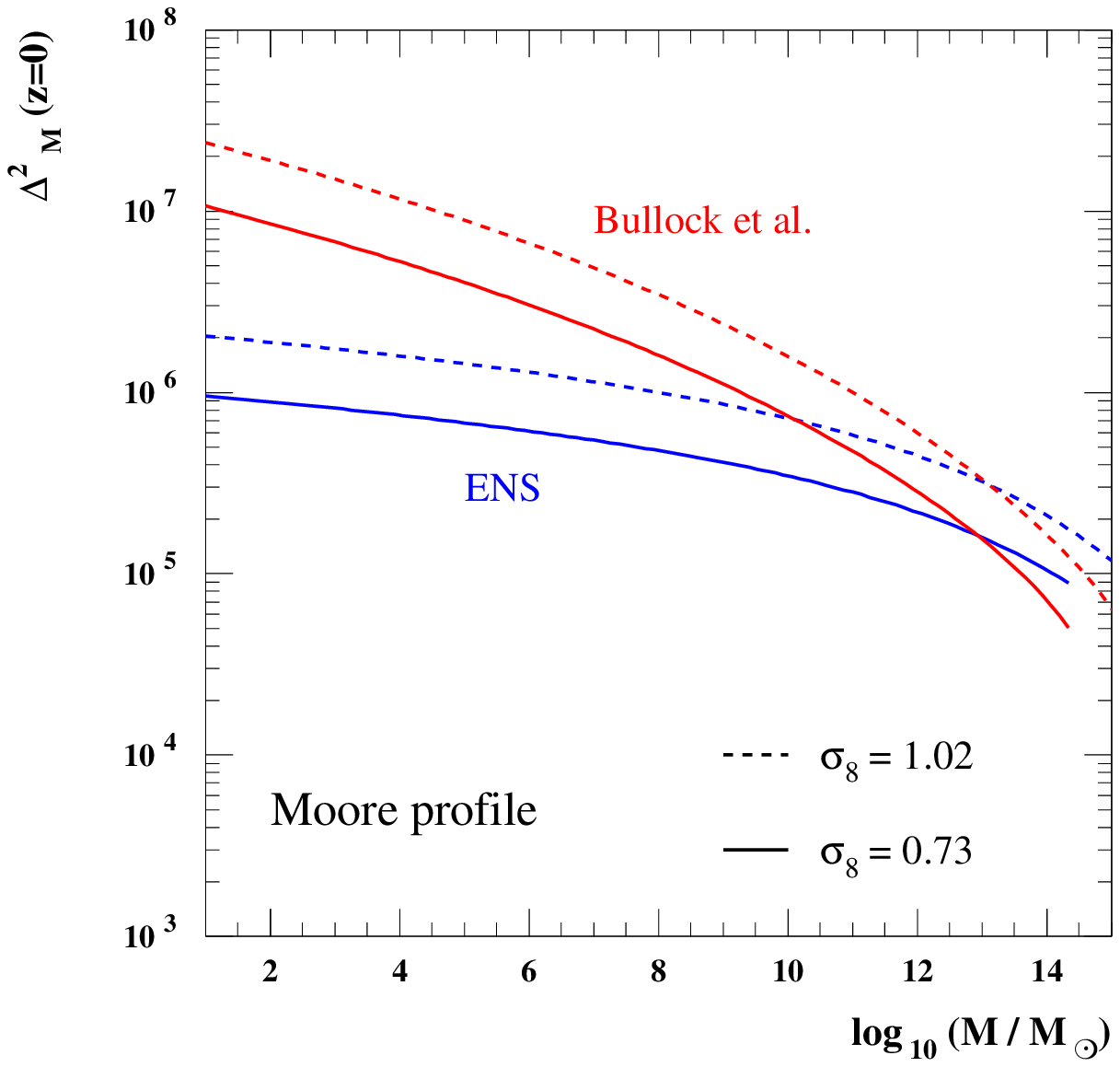}
\includegraphics[width=0.49\textwidth]{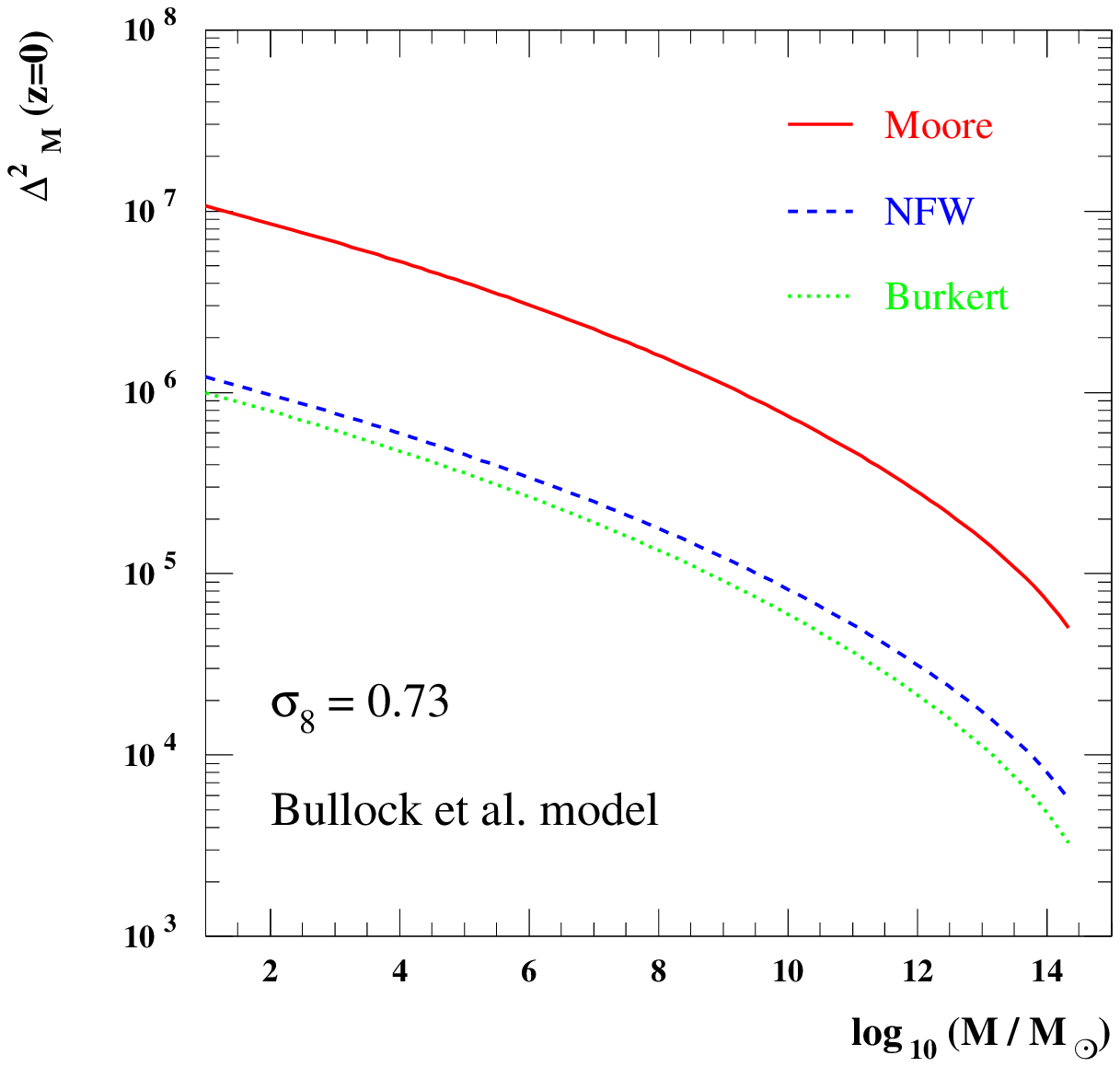}} 
\caption{Average enhancement in the $\gamma$-ray flux emitted in a halo
of mass $M$ at redshift $z=0$ with respect to the case in which the 
same amount of dark matter is smoothly distributed. On the left-hand
side we show how sensitive the result is to the concentration parameter.
On the right-hand side the result for three different families of dark 
matter density profiles is shown.}
\label{fig:fig3}
\end{figure}

\subsection{Flux normalization}

We analyze first how sensitive the flux is to the dark halo properties
we discussed in Section~\ref{sec:halo}. In Fig.~\ref{fig:fig3} -- left 
panel -- we plot $\Delta_M^2$ as a function
of $M$, at redshift $z=0$ and assuming the Moore density profile
to describe dark matter halos. The four cases displayed correspond to the 
two toy models for $c_{vir}$ we have discussed in the previous Section,
and to two choices of $\sigma_8$: our default value 
$\sigma_8 = 0.73$~\cite{2df} and the larger value $\sigma_8 = 1.02$ 
found in another recent analysis~\cite{pier} and more on the line
with values often quoted in the past. For small $M$, i.e.\ large
$c_{vir}$, $\Delta_M^2$ scales roughly like
$c_{vir}^3 / \log^{2}(c_{vir})$, where the logarithmic term follows
from the fact that the halo profiles we are considering have 
logarithmically divergent masses (which we cut at the virial radius).
It follows that the uncertainty on $\sigma_8$ induces about a factor
of 2 uncertainty on $\Delta_M^2$, while an indetermination of a 
factor of a few is due to the model applied to extrapolate $c_{vir}$
to small masses. 

\begin{figure}[t]
\centerline{
\includegraphics[width=0.49\textwidth]{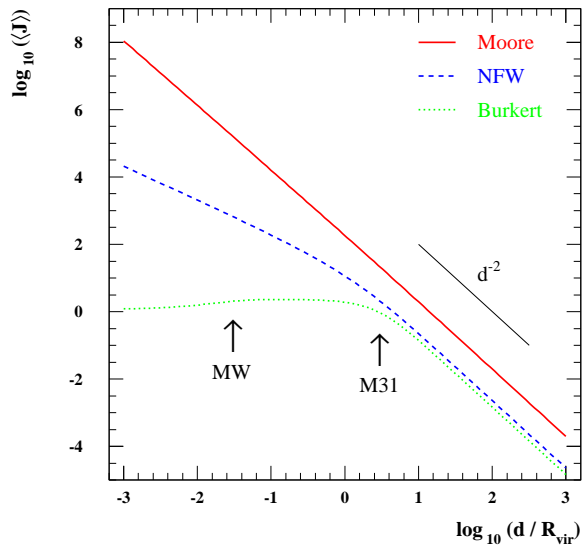}}
\caption{Scaling of the collected $\gamma$-ray flux with the distance $d$ 
between the detector and the center of a halo, for three different halo 
profiles. The angular acceptance of the detector is assumed to be 
$\Delta\Omega = 10^{-3} \rm{sr}$. The plot is for a $10^{12} \Msun$ halo,
the arrows indicate the position on the horizontal axis for the Milky Way
and Andromeda; the case for other masses is analogous.
}
\label{fig:fig4}
\end{figure}

In Fig.~\ref{fig:fig3} -- right panel -- we restrict
to $c_{vir}$ as computed with the Bullock
et al.\ toy model, and show the dependence of the signal on the choice
of halo profile. 
The spread in the predictions between the Moore profile
and the Burkert profile is around a factor of ten independent of mass, 
which is much smaller than the uncertainty due to the choice of profile 
when considering the dark matter induced $\gamma$-ray 
flux generated in single resolved sources. This is one 
of the advantages of considering the cosmological signal. Of course, one is
also less sensitive to the actual halo properties of a single galaxy, 
the Milky Way, which are poorly known. 
This issue is analyzed further 
in Fig.~\ref{fig:fig4} where, for a given halo of density profile 
$\rho(\vec{r})$, we plot the dimensionless quantity
\beq
  \langle\,J\left(\psi\right)\rangle_{\Delta\Omega} 
  = \frac{1}{\Delta\Omega} \frac{1}{8.5\, \rm{kpc}} 
  \cdot \left(\frac{1}{0.3\,{\rm GeV}/{\rm cm}^3}\right)^2
  \int_{\Delta\Omega} d\,\Omega  \int_{l.o.s.} d\,l \;
  \rho^2(\vec{r})\;,
\label{eq:jpsi}
\eeq
a sum over contributions along the line of
sight in a cone of aperture $\Delta\Omega$ in the direction $\psi$
(this quantity often appears in analyses of the WIMP induced flux
generated in the Milky Way halo; normalization factors are fixed
following the choice in Ref.~\cite{bub}). We focus on a
$10^{12} \Msun$ halo, i.e.\ a halo of the size of the Milky Way or 
Andromeda, and assign to it the mean $c_{vir}$ in the Bullock et al 
model; also, we choose $\psi$ in 
the direction of the center of the halo and consider a moderately large 
acceptance angle, $\Delta\Omega = 10^{-3}~\rm{sr}$. We let then the 
distance $d$ between the center of the halo and us vary between 
$10^{-3} R_{vir}$ and $10^{3} R_{vir}$ ($R_{vir} \sim 260\, \rm{kpc}$ 
in our sample case) and plot the corresponding $\langle\,J\,\rangle$ for 
the three halo profiles
we introduced. The arrows in the figure marks the location on the
horizontal axis of the Milky Way (MW) and Andromeda (M31). 
At large $d/R_{vir}$
we find for all halo profiles the $1/d^2$ scaling one expects
for point-like sources: this is obvious for ratios larger than 
$d/R_{vir} \simeq \sqrt{\pi/\Delta\Omega} \simeq 56$, when the halo 
is fully contained in the field of view; however, as it can be seen, 
for the Burkert and the NFW profiles such scaling appears already 
for ratios one order of magnitude smaller, and it is present essentially 
over the whole range displayed for the Moore profile. This
indicates that the bulk of the flux is emitted in the inner halos: 
for the Moore profile 50\% (10\%) of the total emitted flux is generated within
a radius that is about $9\cdot 10^{-6} R_{vir}$ ($6\cdot 10^{-9} R_{vir}$),
for the NFW and Burkert profiles the corresponding radii are shifted, respectively, to
$2.4\cdot 10^{-2} R_{vir}$ ($3.3\cdot 10^{-3} R_{vir}$) and 
$6\cdot 10^{-2} R_{vir}$ ($2.4\cdot 10^{-2} R_{vir}$).
While the spread in predictions for the flux generated in the center of 
our Galaxy is very large (6 orders of magnitude), the total emitted flux
is a much weaker function of the density profile -- the uncertainty is
roughly an order of magnitude. 

This factor of 10 uncertainty is nearly independent of $M$, 
therefore it propagates 
as an order of magnitude uncertainty on the overall normalization 
of the WIMP induced $\gamma$-ray flux.

\begin{figure}[t]
\centerline{
\includegraphics[width=0.49\textwidth]{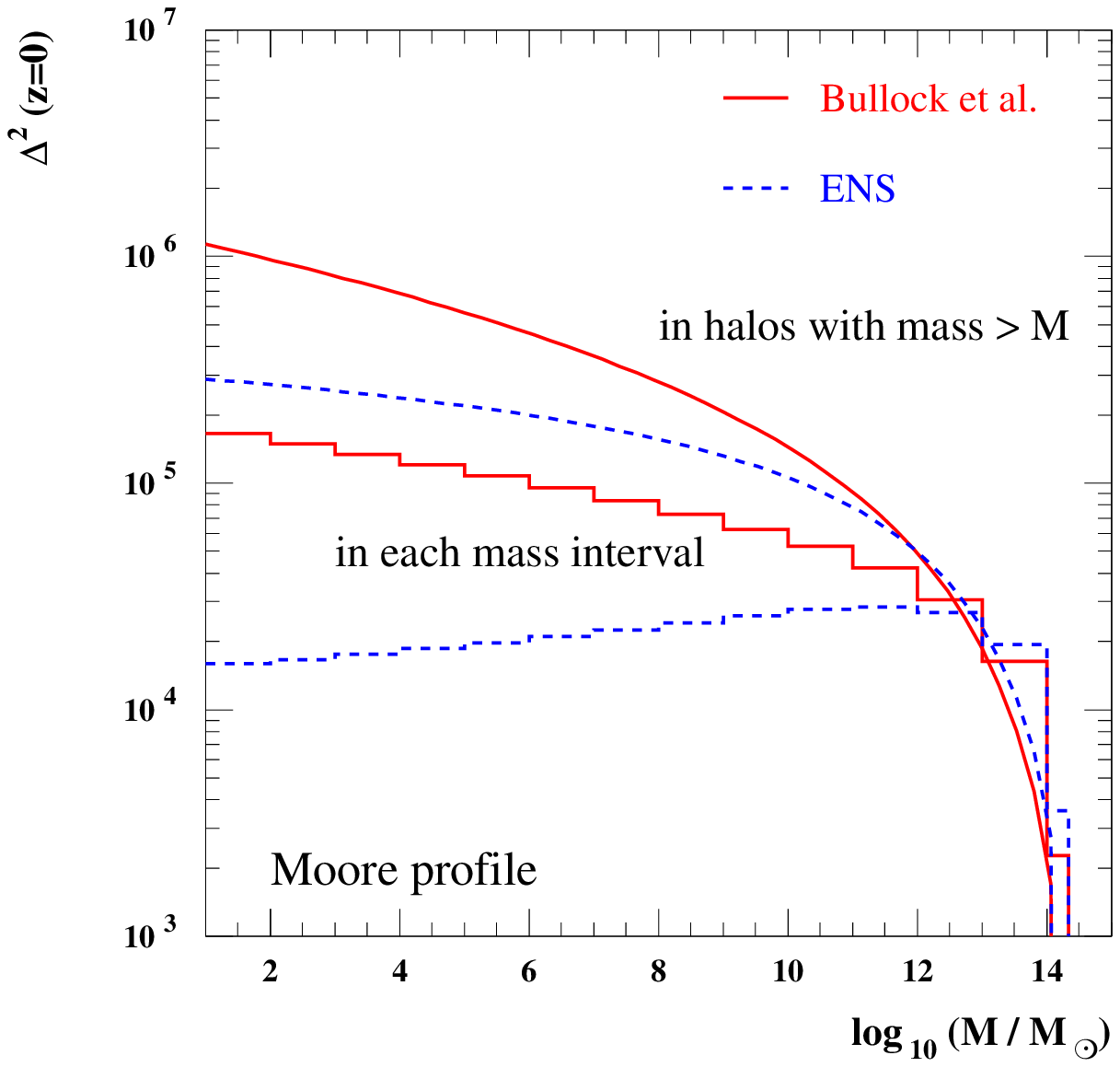}
\includegraphics[width=0.49\textwidth]{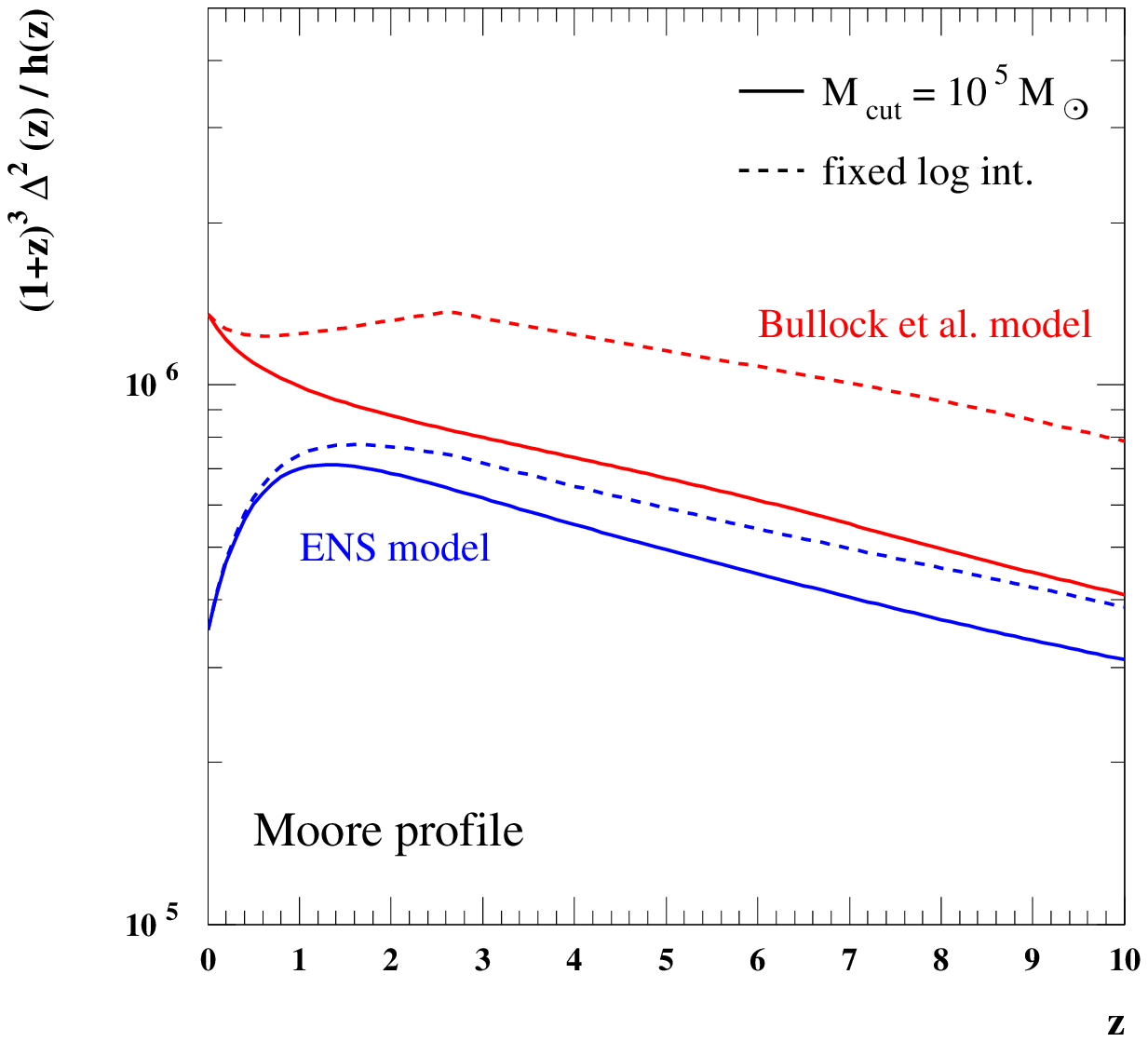}}
\caption{Enhancement in the diffuse $\gamma$-ray flux compared to the case
when all structures in the Universe are erased. On the left-hand side the 
contributions of structures of given masses at $z = 0$ are shown; on the 
right-hand panel we show the redshift dependence, rescaled with
the term $(1+z)^3/h(z)$. 
}
\label{fig:fig5}
\end{figure}

The behavior of $\Delta^2$ is obtained by
folding the scaling of the integrated mass function in 
Fig.~\ref{fig:fig1} with that of $\Delta_M^2$ in Fig.~\ref{fig:fig3}.
The dominant contribution to $\Delta^2$ comes from very small halos:
the integrand in $\Delta^2$ is the product of two mildly divergent 
quantities, the mass function times $M$ and $\Delta_M^2$; the result is 
still convergent but heavily relies on our understanding of the light mass 
end. This is shown in the left-hand panel of Fig.~\ref{fig:fig5}, where,
for the Moore profile and our preferred cosmology, we plot $\Delta^2$ at 
$z=0$ restricting the range of integration over mass. For 
the Bullock et al.\ toy model the contribution per logarithmic interval 
keeps increasing even for the lightest mass range displayed, while in the 
ENS model it starts decreasing but rather slowly. 
Extrapolations of $c_{vir}$ with our toy models to exceedingly small masses 
may not be fully reliable; we prefer to introduce a cutoff in $c_{vir}$
and hence in $\Delta_M^2$ at the
intermediate mass range $M_{cut}$, say $10^5 \Msun$ for $z=0$, where we 
believe the toy models are sufficiently trustworthy. We assume:
\beq 
  c_{vir}(M,z) = c_{vir}(M_{cut},z) \;\;\;\;\;\; \forall \;\; M<M_{cut}
\label{cutoff} 
\eeq 
The choice of $M_{cut}$ is 
to some extent arbitrary; should one make a different assumption
Fig.~\ref{fig:fig5} allows to scale our final results.

In Fig.~\ref{fig:fig5} - right-hand panel, we plot $(1+z)^3 \Delta^2/h$, 
i.e.\ the quantity we need to integrate over $z$
to get the $\gamma$-ray flux, see Eq.~(\ref{eq:flux2}), once
folded with the emission spectrum and the absorption factor. 
We consider both models for $c_{vir}$ and two schemes to define $M_{cut}$. 
In the first we fix $M_{cut}= 10^5 \Msun$ for any $z$, progressively
reducing the mass range over which $c_{vir}$ is extrapolated. Another 
possibility is to keep the range of this extrapolation fixed: at z=0
we choose $M_{\star}/M_{cut}$, with $M_{\star}$ the largest scale allowed
defined implicitly by $\sigma\left(M_{\star}(z)\right)= \delta_{sc}(z)$
and again $M_{cut}= 10^5 \Msun$; at other $z$ the same
ratio is imposed (we never include extrapolations to masses lower than 
$10 \Msun$; at the redshift of a few when $M_{cut}$ would be lower than 
that, we set $M_{cut} = 10 \Msun$). Both schemes are rather arbitrary,
we will show however that the final result is not very sensitive to them.
Notice, on the other hand, the sharp increase of $(1+z)^3 \Delta^2$ at small 
$z$ for the ENS model, whereas a mild decrease or a flat behavior is found
in the Bullock et al.\ model. At larger $z$, the scaling in $1/h(z)$ 
takes over.

\subsection{Spectral signatures}
\label{subsec:spec}

\begin{figure}[t]
\centerline{
\includegraphics[width=0.49\textwidth]{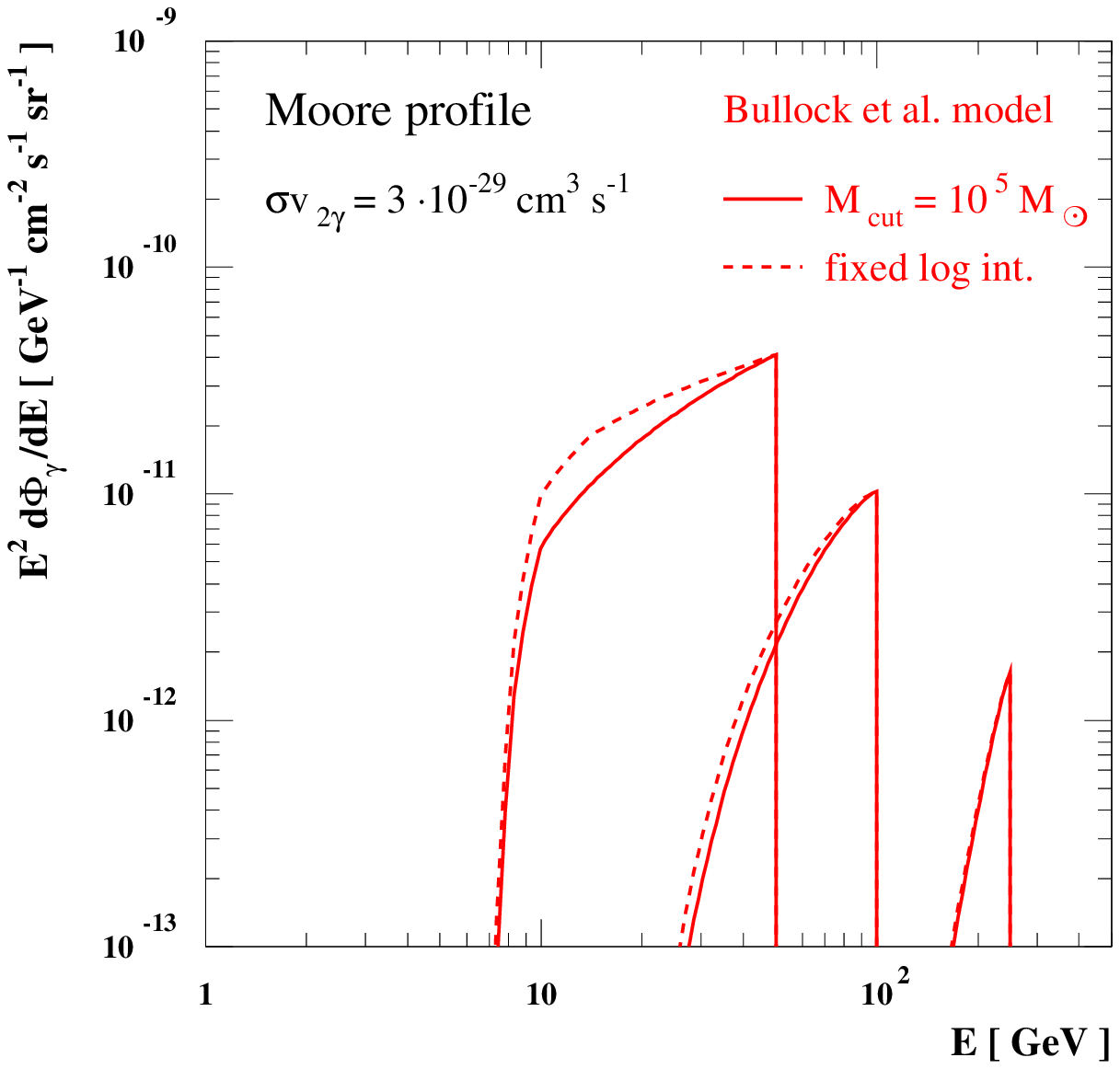}
\includegraphics[width=0.49\textwidth]{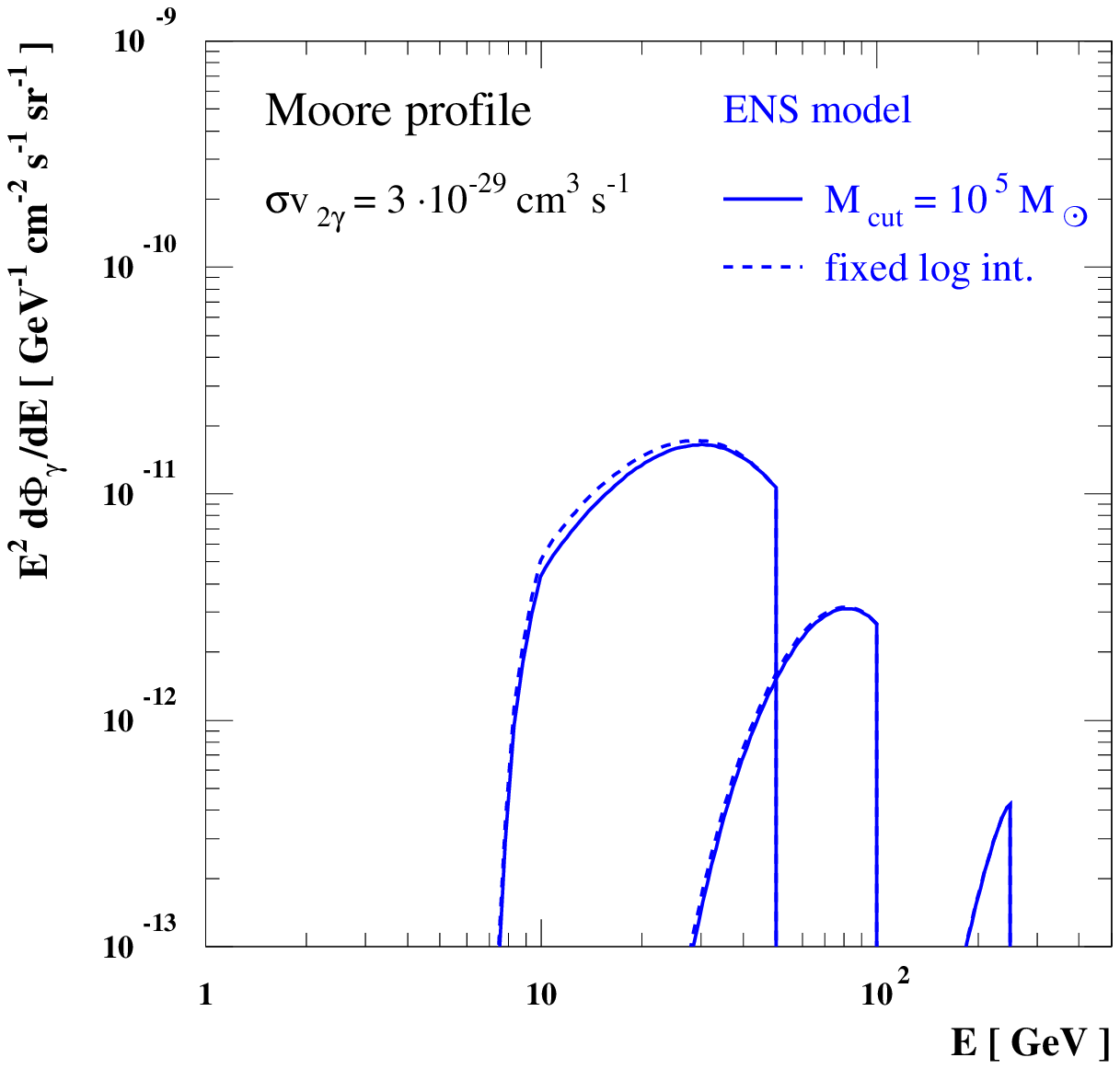}} 
\caption{Spectral signature in the extragalactic gamma-ray flux due to
the annihilation of dark matter WIMPs into monochromatic photons. A toy 
model with three choices of WIMP masses, $M_{\chi} = 50, 100, 250$~GeV,
and fixed annihilation rate into $2\gamma$, is considered. The signature
arises because of the asymmetric distortion of the line due to the 
cosmological redshift, as well as, by absorption of gamma-rays generated 
in distant sources. The normalization of the fluxes
are computed assuming halos are modelled by the Moore profile and 
concentration parameters are derived with the Bullock et al.\ toy model
(left panel) or the ENS model (right panel); solid and dashed curves
refers to two schemes for the choice of the halo mass cutoff $M_{cut}$,
which, as can be seen, plays a marginal role.} 
\label{fig:fig7}
\end{figure}

We now try to estimate, in an approximate way, the level and spectral shape
of the gamma-ray flux that can be expected for a general WIMP, leaving
a more detailed discussion of the extragalactic background one has to
fight against for Section~\ref{sec:bkg}, 
and predicted signals for a more specific (supersymmetric) dark matter 
candidate for
Section\,~\ref{sec:applications}.

The differential gamma-ray yield per WIMP pair annihilation can be written as:
\beq
\frac{dN_\gamma(E)}{dE}= \sum_X b_{\gamma X} n_{\gamma X} 
\delta\left(E-M_{\chi}\,(1-M_X^2/4\,M_{\chi}^2)\right)
+ \sum_F b_{F} \frac{dN_{\rm cont}^F}{dE}(E)\;.
\label{eq:edist}
\eeq
The first term refers to prompt annihilation into two-body final  
states containing a photon, which, forbidden at tree-level essentially
by definition of dark matter (zero electric charge), are allowed at higher 
order
in perturbation theory. Although subdominant, they have the peculiarity
of giving monochromatic $\gamma$-rays: as WIMPs in halos are non-relativistic 
the energy of the outgoing photon is fixed by the WIMP mass $M_{\chi}$ and 
the mass of the particle $X$ (i.e., $E = M_{\chi}$ for the 
$2 \gamma$ final state and $E=M_{\chi}(1-M_X^2/4\,M_{\chi}^2)$ 
for final states with some non-zero mass particle $X$). 
The parameter $b_{\gamma X}$ is the branching ratio into these channels 
and $n_{\gamma X}$ is the number of photons
per annihilation, i.e.\ 2 for the $2 \gamma$ final state and to 1 for the 
others. The second term in Eq.~(\ref{eq:edist}) is instead the term
due to WIMP annihilations into the full set of tree-level final states $F$, 
containing fermions, gauge or Higgs bosons, whose fragmentation/decay chain 
generates photons; this process gives rise to a continuous energy spectrum.

Although there is some span in the predictions for the photon emission rate 
in different particle physics models, the spectral features of the induced
fluxes are quite generic and can be outlined without referring to
a specific model (in section \ref{sec:applications} below we will discuss 
results for more specific models). We start discussing the
monochromatic terms, focusing to be definite on the process
$\chi \bar{\chi} \rightarrow 2 \gamma$ and picking for reference
some typical value for the annihilation cross section in this channel. 
Consider, e.g., that in the simplest case (no resonances or thresholds near the
kinematically released energy in the annihilation $2M_{\chi}$) 
the WIMP total annihilation rate is fixed by the 
approximate relation \cite{jkg}:
\beq
\sigma v \sim \langle \sigma v \rangle \sim 
\frac{3\cdot 10^{-27} \rm{cm}^3 \rm{s}^{-1}}{\Omega_{\chi} h^2} \sim 
3\cdot 10^{-26} \rm{cm}^3 \rm{s}^{-1}\;,
\label{eq:scal}
\eeq
which shows the order of magnitude scaling between the 
thermally averaged annihilation cross section $\langle \sigma v \rangle$ 
and the WIMP thermal relic abundance $\Omega_{\chi}$. Note that this 
relation is only a rough approximation and that large deviations from it
can appear mainly due to resonances and thresholds. In 
Section \ref{sec:applications} below we will not use this approximate 
relation, but instead calculate the relic density including properly both 
coannihilations, resonances and thresholds. For the current discussion 
though, this approximate relation suffices. 
The annihilation into two photons is a 1-loop process
so, in general, its strength is much smaller than $\sigma v$; we assume,
as a sample case when this channel is relevant, $b_{2 \gamma} = 10^{-3}$.

In Fig.~\ref{fig:fig7} we show the induced extragalactic gamma ray flux
for 3 different values of the WIMP mass, $M_{\chi} = 50, 100, 250$~GeV,
and for the two schemes we have considered to estimate $c_{vir}$.
We consider halos modeled by the Moore profile, with no subhalos (the effects
of the latter
will be discussed in Section~\ref{sec:subhalos}).
The figure illustrates the novel signature, first proposed in BEU,
to identify a WIMP induced component in the measured extragalactic 
gamma-ray background, the sudden drop of the gamma-ray intensity at an 
energy corresponding to the WIMP mass due to the asymmetric
distortion of the line caused by the cosmological redshift.
The energy of the $\gamma$-rays at emission determines whether the 
smearing to lower energies has a sharper or smoother cutoff: for a 
larger $M_{\chi}$ the absorption on the extragalactic 
optical and infrared starlight background becomes more efficient.
Spectra obtained applying the ENS toy-model for $c_{vir}$ are similar
to those derived with the Bullock et al.\ model; the main difference, 
for masses lower 
than about 100~GeV, is  a slight shift of the flux peak to lower 
energies. This effect, due to the sharp increase in $\Delta^2$ shown 
in Fig.~\ref{fig:fig5} in a regime where the absorption factor does not 
rapidly take over, tends to reduce the difference in the flux normalization
one might have foreseen looking at $\Delta^2_M$ alone (in the next 
generation of measurements the energy resolution will probably not be 
better than 10\% or so). Fig.~\ref{fig:fig7} also 
illustrates the fact that, at least for the line contributions, the treatment 
of the cutoff in halo mass is not very important; there is
mainly an overall scaling with the choice of $M_{cut}$ at $z=0$, which
the reader can infer from the left-hand panel of Fig.~\ref{fig:fig5}.

\begin{figure}[t]
\centerline{
\includegraphics[width=0.49\textwidth]{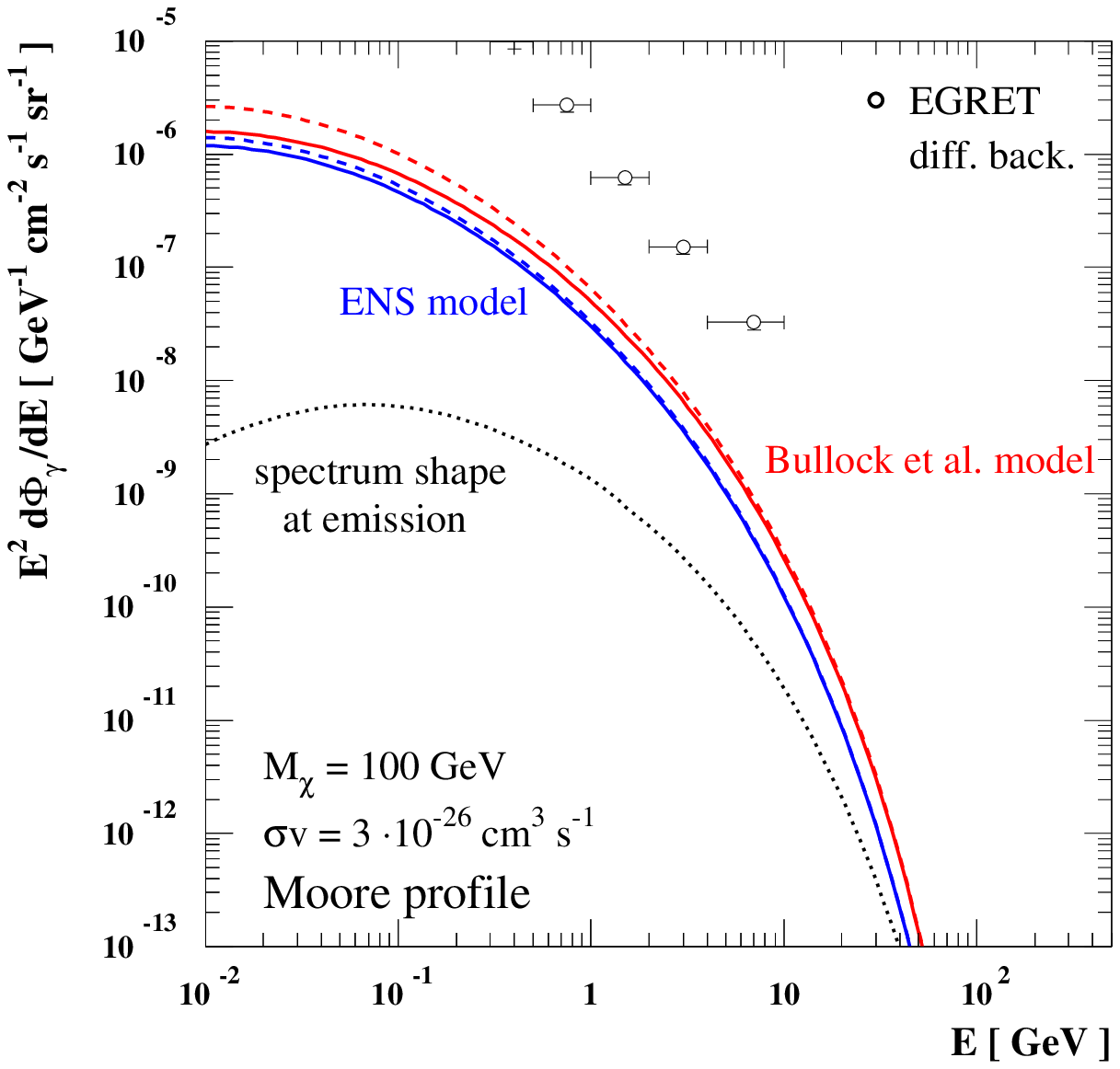}
\includegraphics[width=0.49\textwidth]{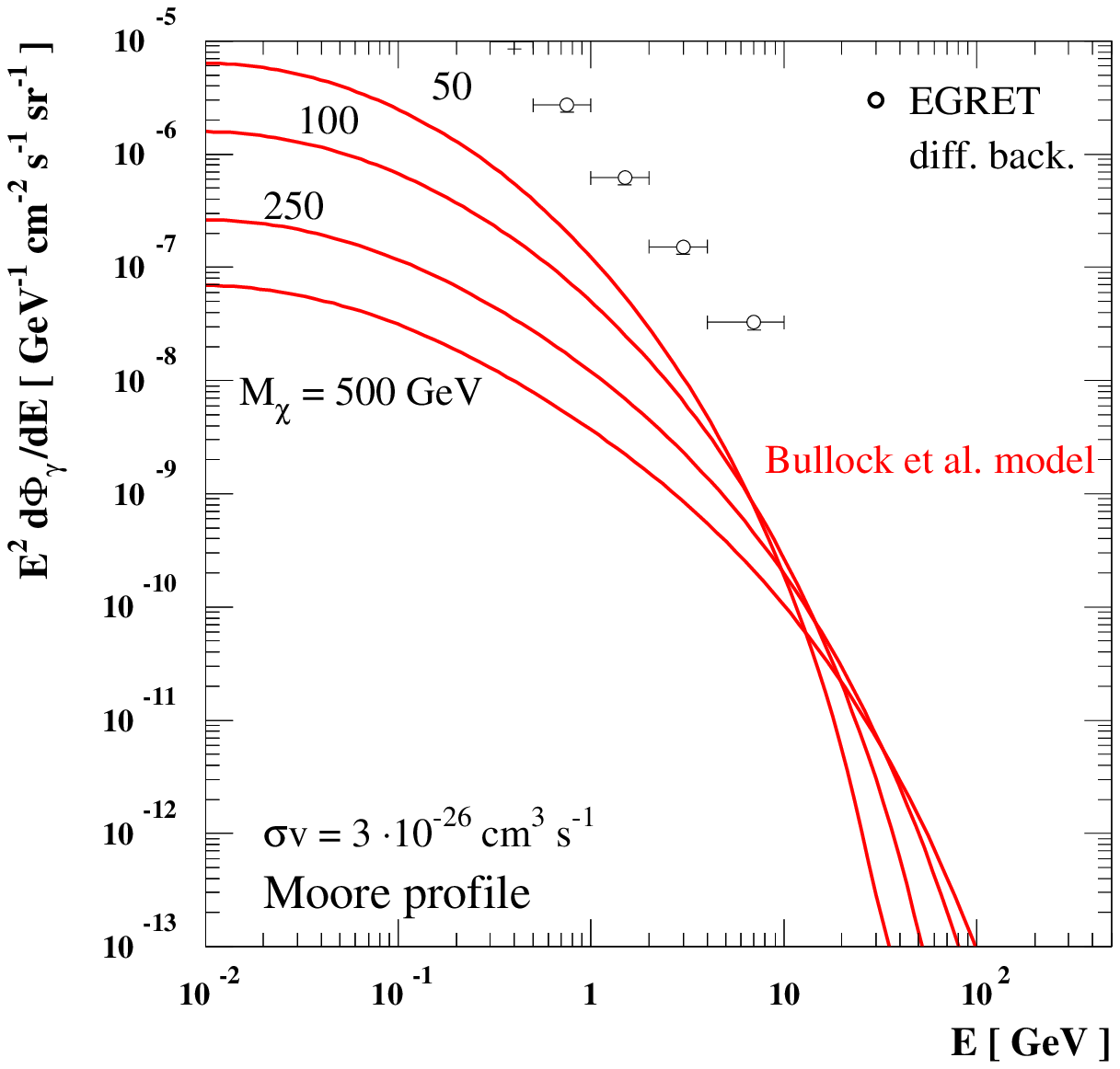}} 
\caption{Spectral features for the extragalactic gamma-ray flux due to
the photons with continuum energy spectrum emitted in pair annihilations
of dark matter WIMPs. The cases considered here are for a WIMP toy model 
of given mass and fixed total annihilation rate, assuming the dominant
branching ratio is $b\bar{b}$. The flux normalizations are computed
under the same assumptions as in Fig.~\protect{\ref{fig:fig7}}. 
In the left panel we compare the shape of the induced flux at Earth
with the one at emission; in the right panel we show the dependence
of the spectral shape on WIMP mass. For comparison, the EGRET estimate
of the extragalactic background flux is shown.} 
\label{fig:fig8}
\end{figure}

Typical features in the continuum contribution are illustrated in 
Fig.~\ref{fig:fig8}. We have assumed again that
$\sigma v \sim 3 \cdot 10^{-26} \rm{cm}^3\,\rm{s}^{-1}$
and supposed that, as often happens in real particle physics models,
the dominant annihilation channel is into $b\bar{b}$;
the energy distribution per annihilating pair in their rest frame 
is simulated with the Pythia Monte Carlo package~\cite{pythia}. 
As most photons are produced in the hadronization and 
decay of $\pi^0$s (98.8\% decay mode: $\pi^0 \rightarrow 2 \gamma$), 
the shape of the photon spectrum is peaked at half the 
mass of the pion, about 70 MeV, and is symmetric around it on a 
logarithmic scale
(sometime this feature is called the ``$\pi^0$ bump'', see, 
e.g.,~\cite{gaisser}).
Had we chosen a different dominant annihilation channel or a combination
of several channels, we would have found very similar behaviors.
As absorption becomes negligible going to low energies,
two features arise when summing extragalactic contributions
over all redshifts: The peak in the spectrum is shifted to lower energies
and there is a sharper decrease in the flux approaching the 
value of the WIMP mass. The first signature is probably
hidden in background fluxes, see the discussion in the next Section.
The second feature is instead potentially interesting, especially in case
the line components are negligible: 
While a sensible contribution to the extragalactic 
$\gamma$-ray background can be provided by WIMPs in the few GeV energy 
range, at higher energies the WIMP induced flux is very rapidly suppressed.
Such behavior cannot be associated to a spectral index, while
background components are closer to a power law.

As shown on the right-hand side of Fig.~\ref{fig:fig8}, the WIMP-induced
extragalactic flux gradually flattens for heavier and heavier WIMPs;
also shown is the current estimate of the diffuse extragalactic background flux
as derived from the analysis of data taken by the EGRET 
telescope~\cite{Sreekumar98}.

\subsection{Role of subhalos}\label{sec:subhalos}

We have shown that small dense halos are providing the bulk of the WIMP 
induced $\gamma$-ray flux. So far we have considered just the case for
isolated halos; as already mentioned, N-body simulations indicate that, 
in the clustering process with large halos forming by the merging of 
smaller objects, a fraction of the latter, up to about 10\% of the total mass,
may have survived tidal disruption and appear as bound subhalos inside 
virialized halos~\cite{ghigna,klypin99}. 
From the point of view of structure formation, the presence of rich 
substructure populations was at first seen as the main flaw in the picture 
from N-body simulations of $\Lambda$CDM cosmologies, a ``crisis'' urging 
for a solution~\cite{ghigna,klypin99}, maybe with a drastic change in the 
particle physics set up, see, e.g.,~\cite{da-ho}. More recent analyses 
indicate that those results should be reinterpreted and the apparent 
discrepancy 
between the number of subhalos found in the simulations and that of luminous 
satellites observed in real galaxies is fading away, see, 
e.g.,~\cite{hayashi,stoehr,bullocketal,bensonetal}.
From the point of view of dark matter detection, substructure may play a 
crucial role~\cite{silkstebbins}, even in the interpretation of
currently available data. 
Consider, e.g., the gamma-ray halo surrounding the Galaxy for which
statistical evidence has been claimed in data collected by the EGRET 
telescope~\cite{dixon}: the conjecture that this may be generated by pair
annihilations of relic dark matter particles is based on the possible
presence of dense dark matter clumps in the Milky Way halo~\cite{grhalo}.

At first sight, the role of substructure may seem marginal in our 
context. The fraction of mass in subhalos is small and the subhalo mass 
function is not likely to be significantly steeper than the mass function
for isolated halos, Eq.~(\ref{eq:massfunc}). We expect then the number of halos
in a given mass range to be larger than the number of substructures in the 
same range. On the other hand, the concentration parameter in subhalos may 
be significantly larger than for halos: on average, subhalos arise in 
higher density environments, as well as we expect a depletion in their 
outskirts by tidal stripping. This trend has indeed been observed in the 
numerical simulation of Ref.~\cite{Bullock}, where it is shown
that, on average and for $M \sim 5\cdot 10^{11} \Msun$ objects, the 
concentration 
parameter in subhalos is about a factor of 1.5 larger than for halos. 

Consider a halo of mass $M$ and suppose that, on average, a fraction $f$ of 
its total mass is provided by substructures with mass function $dn_s/dM_s$. 
The differential energy spectrum for the number of photons emitted in such a
halo, rather than by Eq.~(\ref{eq:dnde}), is now given by:
\beqa
  \frac{d{\cal N}_{\gamma}}{dE} (E,M,z) & = & \frac{\sigma v}{2}
  \frac{dN_{\gamma}(E)}{dE} \left[ \int dc^{\,\prime}_{vir}\; 
  {\cal{P}}(c^{\,\prime}_{vir})
  \left(\frac{(1-f)\,\rho^{\prime}}{M_{\chi}}\right)^2 
  \int d^3r\; g^2(r/a) \right. \nonumber \\
  && \left. + \int dM_s \frac{dn_s}{dM_s} \int dc^{\,\prime}_{vir}\; 
  {\cal{P}}_s(c^{\,\prime}_{vir})
  \left(\frac{\rho^{\prime}}{M_{\chi}}\right)^2 
  \int d^3r\; g^2(r/a) \right] \nonumber \\
  &=& \frac{\sigma v}{2} \frac{dN_{\gamma}(E)}{dE}   
  \frac{\bar{\rho}_c}{M_{\chi}^2} \left[(1-f)^2 M \Delta_M^2(z,M)
  + \int dM_s \frac{dn_s}{dM_s} M_s \Delta_{M_s}^2(z,M_s) \right]
\label{eq:dndenew}
\eeqa
A simple ansatz is that the subhalo mass function has a power-law
behavior $dn_s/dM_s \propto 1/M_s^{\beta}$ for $M_s < M$
($\beta < 2\ $ is required 
for the total mass to be finite), with the normalization fixed by using the 
definition of $f$, i.e.
\beq
   \int dM_s M_s \frac{dn_s}{dM_s} = f M\;.
\eeq
This gives
\beq
   \frac{dn_s}{dM_s} = (2-\beta) f \frac{M^{\beta-1}}{M_s^{\beta}}\;.
\label{eq:submass}
\eeq
If we further assume that $f$ and $dn_s/dM_s$ are independent of $M$,
inserting Eq.~(\ref{eq:submass}) into Eq.~(\ref{eq:dndenew}), we find 
that the contribution of subhalos can be included in the formula for 
the gamma-ray flux, Eq.~(\ref{eq:flux2}), with the replacement:
\beq
  \Delta_M^2(z,M) \; \rightarrow \; (1-f)^2 \; \Delta_M^2(z,M) +
  (2-\beta) f M^{\beta-2} \int dM_s \,M_s^{1-\beta}\, 
  \Delta_{M_s}^2(z,M_s)\;.
\label{eq:D2Ms}
\eeq
Here $\Delta_{M_s}^2$ is just $\Delta_{M}^2$ but with values of $c_{vir}$
and ${\cal{P}}(c^{\,\prime}_{vir})$ appropriate for the subhalos.
It may be premature to deduce the latter from N-body 
simulation results.
The scaling $c_{vir} \propto M^{-0.3}$ proposed in Ref.~\cite{Bullock} 
probably cannot be extrapolated to small masses: for $10^{5} \Msun$ 
subhalos we would get a value of the concentration parameter 40 times 
larger than the value for halos as computed with the Bullock et al.\ toy 
model. 

A prediction for the subhalo mass function is missing as well; 
there are just limited studies, not fully covering the mass range we are 
interested in. The current N-body simulation results are consistent 
with a power law behavior, but with non-universal slope and some 
indication that the index $\beta$ is getting harder decreasing the mass 
of the host halo. We find, e.g., from Fig.~5 in Ref.~\cite{springel} that 
$\beta \simeq 1.66$ for a $10^{15} \Msun$ halo. A few studies are
focused on Milky Way size halos, $10^{12} \Msun$: from, e.g., Fig.~1 
in Ref.~\cite{stoehr} we can extract the scaling 
$dn_s/dM_s \propto 1/M_s^{1.95}$. 

The value of $f$ is a matter of debate as well. Values for the fraction
of mass in substructures quoted in the literature are in the range
1 \% to 10 \% and they often refer to the ratio of the sum of the masses of 
identified subhalos to the total mass, rather than to an extrapolation
performed assuming a mass function. Such a value, say $f^{\prime}$, should 
then depend both on the algorithm for finding subhalos in the simulation, 
and, most importantly, on the resolution of the simulation. Suppose
that $f^{\prime}$ refers to a simulation where, for halos of mass $M$,
substructures of mass down to $p\,M$ can be resolved; then, with our 
notation,
\beq
   f^{\prime} M = f M^{\beta-1} \left[ M^{2-\beta}- (p\,M)^{2-\beta}\right]
   \,,
\eeq
i.e., $f = f^{\prime} / (1 - p^{2-\beta})$. If $M=10^{12} \Msun$ and 
$p=5 \cdot 10^{-5}$~\cite{stoehr}, we find $f = 2.56 f^{\prime}$.
It is then not implausible that the true $f$, eventually to be found at 
future ultra-high resolution simulations, may approach or even exceed 
10 \%.

\begin{figure}[t]
\centerline{
\includegraphics[width=0.49\textwidth]{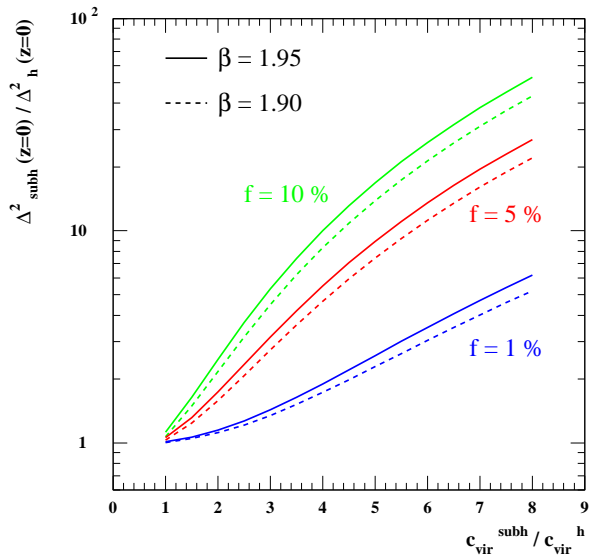}}
\caption{Influence of substructure on the flux normalization for three
different average fractions $f$ of the total mass in subhalos; we have 
restricted to a specific mass function (see text) with spectral index
$\beta$ and kept as a free parameter, we display on the horizontal axis 
the mean enhancement in the concentration parameter in subhalos.
}
\label{fig:fig6}
\end{figure}

To give a feeling for the possible effect of substructure, we consider the
simplified sample case in which $f$ and the mass function are universal, and
keep the average enhancement in the concentration parameter as a free 
parameter (we find the value of $c_{vir}(M,z)$ for subhalos by a rigid
rescaling of the $c_{vir}(M,z)$ found for halos of same mass and at the 
same $z$: actually, the mass range in which this rescaling matters is 
just around the cutoff mass $M_{cut} = 10^5 \Msun$).
In Fig.~\ref{fig:fig6} we consider $\beta = 1.95$ or the slightly softer
$\beta = 1.90$, choose three sample values for the fraction of the mass 
in subhalos $f$ and plot the ratio of the 
value of $\Delta^2$ with and without including subhalos as a function 
of the average enhancement in the concentration parameter.
Sensible gains in $\Delta^2$ and hence in
the $\gamma$-ray flux normalization are viable even for moderate 
enhancements in the concentration parameter. Again, the effect of 
substructure is less dramatic than in case of single dark matter 
sources: the argument here is analogous to the one presented in the 
discussion on the role of the singularity in halo profiles.

\subsection{Observability of subhalos in the Milky Way halo}
\label{subs:mwhalos}

It would be of utmost importance to test the subhalo picture
predicted by CDM N-body simulations by collecting information 
from the morphology of the Milky Way halo. As already mentioned,
a rich population of luminous satellites is not observed 
in the Galaxy and this was considered, up to recent work, one 
the most severe ``problems'' of CDM. There are now 
models~\cite{bullocketal,bensonetal} to explain why small substructures
may be totally dark (without visible baryons); if this is indeed the case,
WIMP annihilation might be the only chance to perform a detailed mapping 
of the distribution of mass in the Milky Way. This issue has been 
investigated by numerous authors (for a recent analysis see, e.g., 
Ref.~\cite{clumps}). The problem however reduces to the study  of the 
actual realization of incalculable random processes and this implies that it
is very hard to estimate the probability for detection of a signal.
In particular, a crucial parameter will be the location of 
the nearest dark matter clump, since this will dominate the signal.
We show here how the picture we have outlined for a generic halo applies
to the Milky Way and discuss the implications for the observability of 
subhalos.

The gamma-ray flux from a single ``clump'' of mass $M_s$ and at the distance
$d$ from the Earth is equal to:
\beqa
   \frac{d\phi_{\gamma}^{1-cl}}{dE} &=& \frac{\sigma v}{2} 
   \frac{dN_{\gamma}(E)}{dE} \frac{1}{4 \pi} 
   \int_{\Delta\Omega} d\,\Omega  \int_{l.o.s.} d\,l \; 
   \left(\frac{\rho(\vec{r})}{M_{\chi}}\right)^2 \nonumber \\
   &=& 9.35 \cdot 10^{-11}
   \left(\frac{\sigma v}{10^{-26}\;\rm{cm}^3 \rm{s}^{-1}}\right)
   \left(\frac{100\;\rm{GeV}}{M_{\chi}}\right)^2 \frac{dN_{\gamma}(E)}{dE}
   \Delta\Omega\; \langle\,J\left(\psi=0\right)\rangle_{\Delta\Omega}\;
   \rm{cm}^{-2} \rm{s}^{-1} \rm{GeV}^{-1}\,.
\eeqa
The angular extension of most clumps is much smaller than $\Delta\Omega$,
hence we can use the point source approximation. In our formalism the 
formula becomes:
\beqa
   \frac{d\phi_{\gamma}^{1-cl}}{dE} &=& \frac{\sigma v}{2} 
   \frac{dN_{\gamma}(E)}{dE} \frac{1}{4 \pi \, d^2}
   \frac{M_s}{M_{\chi}^2} \, \frac{\Delta_{vir}\bar{\rho}_0}{3}\,
   \frac{(c^{\,\prime}_{vir}\,x_{-2})^3}
   {\left[I_1(c^{\,\prime}_{vir}\,x_{-2})\right]^2} \,
   I_2(x_{min},c^{\,\prime}_{vir}\,x_{-2})
\eeqa

\begin{figure}[t]
\centerline{\includegraphics[width=0.49\textwidth]{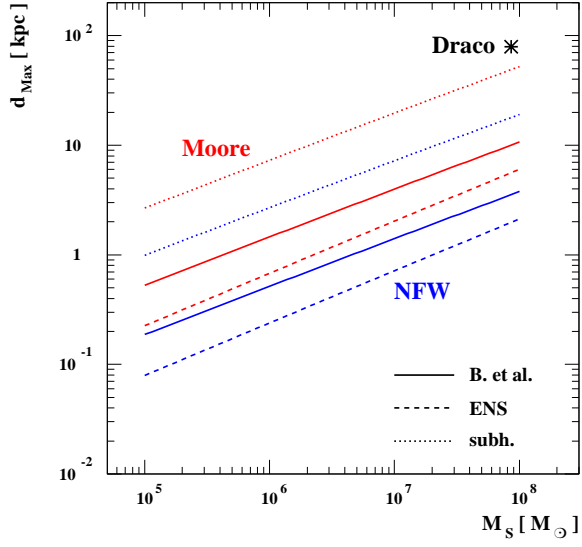}}
\caption{Maximum distance $d_{Max}$ of a clump from the our position in 
the Galaxy for which the $\gamma$-ray flux due to WIMP annihilations in 
the clump exceeds the estimated point source sensitivity of GLAST. We 
picked a specific WIMP dark matter candidate, while we are considering 
a few models to relate the mass of the clump $M_s$ to its concentration 
parameter, as well as two models for the halo density profiles in the 
clump (see the text for details)}
\label{fig:point}
\end{figure}

Assume $\phi_{\gamma}(E>100\;\rm{MeV}) > 1.6 \cdot 10^{-9} \rm{cm}^{-2} 
\rm{s}^{-1}$~\cite{Ritz-glast} is the point source sensitivity of the next 
gamma-ray telescope in space, GLAST, which, as EGRET did, will map the whole 
gamma-ray sky. In defining a particle physics model for a WIMP, one has to
fix $M_{\chi}$, $\sigma v$ and the branching ratios into each annihilation 
channel. It is then possible to compute, say, 
\beq
N_{\gamma}^{100} = \int_{100\;\rm{MeV}} dE^{\prime} 
\frac{dN_{\gamma}(E^{\prime})}{dE^{\prime}}\,. 
\eeq
For each $M_s$ we can estimate the maximum distance $d_{Max}$ of the 
clump from us such that the WIMP induced flux is larger than the point source 
sensitivity of GLAST. This is shown in Fig.~\ref{fig:point} for one of the 
WIMP toy models introduced in Ref.\ \ref{subsec:spec}: we assume 
that $M_{\chi} = 50\,\rm{GeV}$ and that the total annihilation rate into
$b\bar{b}$ is $\sigma v = 10^{-26}\;\rm{cm}^3 \rm{s}^{-1}$, and find
$N_{\gamma}^{100} = 25.9$; a generalization to other models can be obtained 
very simply by scaling of these values.
Six configurations for the normalization of the flux 
are chosen: we assume that $c_{vir}(M,z=0)$ for subhalos is either equal
to the mean value found with the Bullock et al.\ or ENS toy models for 
isolated 
halos (labels ``B. et al.'' and ``ENS'' respectively) or to 4 times the
value found with the Bullock et al.\ model (label ``subh.''); we
consider also the cases that subhalos are described both by the Moore et al.\ 
profile and by the NFW profile (the results for the Burkert profile are very 
close to the ones for NFW, unless the clump is very close to us, see 
Fig.~\ref{fig:fig4}). As can be seen, unfortunately, the prediction 
of our analysis is that just a few nearby clumps might be detected by GLAST.
For comparison, we show the location in the plane of the figure of
a ``clump'' that is sufficiently massive to have a luminous counterpart, 
Draco. This dwarf spheroidal, together with other similar candidates, 
has been considered several times in the literature as a potential 
gamma-ray dark matter source, see, e.g.,~\cite{tyler} (note, however,
that our picture applies on average, rather than to a single specific 
source, which might be better characterized through its rotation curve).

\section{The diffuse extragalactic gamma-ray background}\label{sec:bkg}

The observability of the signal proposed here depends on the level of the 
diffuse extragalactic gamma ray background. Contributions from several 
classes of unresolved discrete sources have been discussed in the literature. 
After EGRET maps of the gamma-ray sky, the case for a dominant contribution
from blazars is generally considered to be very strong: a large number of
high-energy emitting blazars has been observed and, as we will show
and contrary to other candidates, their distribution of energy spectra 
seems to be compatible with the observed extragalactic radiation. We will 
then rederive here the expected diffuse background under the assumption that 
the source of the background is unresolved blazars. We will mostly follow 
the analysis of Salamon and Stecker \cite{StSaMa93,SaSt94,StSa96}, 
but update it with more recent data and examine the expected uncertainties.

\subsection{Basic blazar model}

The basic model assumes that the diffuse gamma ray flux comes from unresolved blazars. We will assume that the blazars are distributed in redshift and luminosity according to a luminosity function $\rho_\gamma(P_\gamma,z)$ where $P_\gamma$ is the luminosity (in units of W Hz$^{-1}$ sr$^{-1}$). The luminosity function $\rho_\gamma$ is the comoving density in units of Mpc$^{-3}$ (unit interval of $\log_{10} P_{\gamma/r}$)$^{-1}$\@. We will further assume that the blazars emit gamma rays with some spectral index $\alpha$ which is distributed according to a distribution function $p(\alpha)$. The absorption of gamma rays emitted at redshift $z$ and observed at energy $E_0$ is, as before, parameterized in terms of the optical depth $\tau(E_0,z)$ such that the attenuation is proportional to $e^{-\tau(E_0,z)}$. We will here use the parameterization of the Kennicut model in Primack et al.\ \cite{abs} introduced in Section \ref{sec:dm}. For comparison we will also use the model of Salamon and Stecker \cite{SaSt98} (their Fig.~6 with metallicity correction). There is also a recent estimate of the absorption by de~Jager and Stecker \cite{deJaSt02}, but we will not use that model since it is not valid above $z\simeq0.3$ which is not sufficient for our purpose. 
In EGRET observations one has seen that blazars most of the time are in a quiescent state but some small fraction of time are in a flaring state with higher luminosity and slightly different spectral index (softer, i.e.\ higher $\alpha$). We will assume that the blazars are in the flaring state a fraction $\zeta$ of their time and that their luminosity then is a factor $A_f$ higher than in the quiescent state. We will also parameterize the different spectral indices by assuming that they come from the same distribution function $p(\alpha)$ but shifted by $\Delta\alpha_q$ and $\Delta\alpha_f$ for the quiescent and flaring states respectively (these two quantities are not independent; we have adopted 
here the same notation as in Salamon and Stecker, but, alternatively, one 
could redefine $\alpha$ and introduce a single shift $\Delta\alpha$).
Putting this together we can write the gamma ray flux (in units of cm$^{-2}$ s$^{-1}$ sr$^{-1}$ GeV$^{-1}$) at energy $E_0$ as
\begin{eqnarray}
  \phi_\gamma(E_0) & = & \frac{c}{H_0}
  \int_{\log_{10} P_\gamma^{\rm min}}^{\log_{10} P_\gamma^{\rm max}} d\log_{10}{P_\gamma}
  \int_{\alpha_{\rm min}}^{\alpha_{\rm max}} d\alpha
  \int_{z_{\rm min}}^{z_{\rm max}} dz \nonumber \\
  & & 
  \frac{1}{h(z)}
  \frac{P_\gamma}{2 \pi \hbar E_{\gamma,f}} \Big[
   p(\alpha-\Delta\alpha_q) \rho_{\gamma,q}(P_\gamma,z)+
   p(\alpha-\Delta\alpha_f) \rho_{\gamma,f}(P_\gamma,z) \Big]
  \frac{dN_\gamma}{dE}(E_0(1+z),\alpha)
  e^{-\tau(E_0,z)}.
  \label{eq:gammabkg}
\end{eqnarray}
In this equation, we have introduced the following
\begin{equation}
  \left\{ \begin{array}{lcl}
  H_0 & = & \mbox{Hubble constant today} \\
  h(z) & = & \mbox{cosmology factor as defined in Eq.~(\ref{eq:hofz})} \\
  E_{\gamma,f}=0.1~\mbox{GeV} & = & \mbox{the fiducial gamma ray energy at which
    the luminosity is $P_\gamma$} \\
  \frac{dN_\gamma}{dE} & = & \mbox{the gamma ray spectrum
    (normalized to 1 at $E_{\gamma,f}$)} \\
  \hbar & = & \mbox{Planck's reduced constant} \\
  \rho_{\gamma,q} & = & \mbox{the luminosity function for blazars in quiescent state} \\
  \rho_{\gamma,f} & = & \mbox{the luminosity function for blazars in flaring state}
  \end{array}
  \right.
\end{equation}
Note that we have for clarity explicitly included $c$ and $\hbar$ in Eq.~(\ref{eq:gammabkg}), but
the unit conversion factors to get the flux in the above given units are not given explicitly. Note that there is no factor of $1/4\pi$ since the luminosity $P_\gamma$ is per sr already. We will, as before, assume that $H_0=70$ km s$^{-1}$ Mpc$^{-1}$, $\Omega_m=0.3$ and $\Omega_\Lambda=0.7$.

In the following subsections we will go through the different factors entering in Eq.~(\ref{eq:gammabkg}).

\subsection{The luminosity function $\rho_\gamma(P_\gamma,z)$}

We need to know the luminosity function as a function of redshift. Since not that many blazars are observed we will follow \cite{StSaMa93,SaSt94,StSa96} and assume that the same basic mechanism (i.e.\ the same population of high-energy electrons) is responsible for both the gamma ray and the radio flux. We can then use the much larger catalogs of radio sources to get the luminosity function. We will assume that the luminosities in gamma and radio are related by
\begin{equation}
  \left\{ \begin{array}{lcl}
  P_{\gamma,q} & = & \kappa P_r \\
  P_{\gamma,f} & = & A_f \kappa P_r
  \end{array} \right.
\end{equation}
where $P_\gamma$ and $P_r$ are the luminosities (in units of W Hz$^{-1}$ sr$^{-1}$). The gamma ray luminosity is given at 0.1 GeV and the radio luminosity at 2.7 GHz. The subscripts $q$ and $f$ refer to the quiescent and flaring states respectively.
We will assume that the two luminosity functions are related by
\begin{equation}
  \rho_\gamma(P_\gamma,z) = \eta \rho_r(P_r,z)
\end{equation}
where $\rho_\gamma$ and $\rho_r$ are the luminosity functions (in units of Mpc$^{-3}$ (unit interval of $\log_{10} P_{\gamma/r}$)$^{-1}$). The factor $\eta$ takes into account possible beaming effects which could mean that not all radio blazars emit gamma rays towards the Earth (or vice versa). Including the effect that the blazars are assumed to be in the flaring state a fraction $\zeta$ of the time, we can finally write
\begin{equation}
  \left\{ \begin{array}{lcl}
  \rho_{\gamma,q}(P_\gamma,z) & = & (1-\zeta) \eta \rho_r(\frac{P_\gamma}{\kappa},z) \\
  \rho_{\gamma,f}(P_\gamma,z) & = & \zeta \eta \rho_r(\frac{P_\gamma}{A_f \kappa},z)
  \end{array} \right.
\end{equation}
For the radio luminosity function, we use the parameterization by Dunlop and Peacock \cite{DunlopPeacock90}
\begin{equation}
  \rho_r(P_r,z) = 10^{-8.15} \left[ 
  \left( \frac{P_r}{P_c(z)} \right)^{0.83}+
  \left( \frac{P_r}{P_c(z)} \right)^{1.96}\right]^{-1}
  \quad ; \quad P_c(z) = 10^{25.26+1.18z-0.28z^2}
  \label{eq:radiolf}
\end{equation}
valid up to $z=5$. This luminosity function was derived for a cosmology with $H_0=50$ km s$^{-1}$ Mpc$^{-1}$ and $\Omega_0=\Omega_m=1$, but we can approximately convert this to a luminosity function for our cosmology by multiplying $\rho$ with a correction factor \cite{DunlopPeacock90}
\begin{equation}
   \frac{dV_{\rm std}}{dV} = \frac{
    \left( \frac{(R_0 S_k(r))^2}{H_0 h(z)} \right)_{\Omega_0=\Omega_m=1,
      H_0=50~{\rm km\, s^{-1}\, Mpc^{-1}}}}
   {\left( \frac{(R_0 S_k(r))^2}{H_0 h(z)} \right)_{\rm our~cosmology}}
\end{equation}
and multiplying $P_c(z)$ with the correction factor
\begin{equation}
\left[
\frac{\left(D_L\right)_{\rm our~cosmology}}
{\left(D_L\right)_{\Omega_0=\Omega_m=1,
      H_0=50~{\rm km\, s^{-1}\, Mpc^{-1}}}}
\right]^2 =
\left[
\frac{\left((1+z)R_0 S_k(r)\right)_{\rm our~cosmology}}
{\left((1+z)R_0 S_k(r)\right)_{\Omega_0=\Omega_m=1,
      H_0=50~{\rm km\, s^{-1}\, Mpc^{-1}}}}
\right]^2
\end{equation}
where $D_L$ is the luminosity distance.
The luminosity function Eq.~(\ref{eq:radiolf}) is valid between $P_r^{\rm min}=10^{18}$ W Hz$^{-1}$ sr$^{-1}$ and $P_r^{\rm max}=10^{30}$ W Hz$^{-1}$ sr$^{-1}$ which we will convert to limits on $P_\gamma$. Note that the exact upper and lower limits on the luminosity are unimportant since $P_\gamma \rho_\gamma$ that enters in Eq.~(\ref{eq:gammabkg}) is peaked well between the lower and upper limits and is vanishingly small at the boundary.
For the redshift integration we will as a default integrate between $z_{\rm min}=0$ and $z_{\rm max}=5$, but this integration range will be, as discussed below, restricted to include the effect of resolved blazars.

For the parameters $\kappa$, $\eta$, $\zeta$ and $A_f$, we will use the values obtained in \cite{StSa96},
\begin{equation}
  \left\{ \begin{array}{lcl}
  \kappa & = & 4 \cdot 10^{-11} \\
  \eta & = & 1.0 \\
  \zeta & = & 0.03 \\
  A_f & = & 5.0 \\
  \end{array} \right.
\end{equation}
where $\kappa$ was determined from observations of blazars that are observed both in radio and in gamma rays, $\eta$ was determined by requiring the number counts of blazars to be consistent with the EGRET observations and $\zeta$ and $A_f$ was determined from EGRET blazar observations.

\subsection{Intrinsic gamma ray spectrum}

We will assume that the intrinsic gamma ray spectrum follows a power law with spectral index $\alpha$, i.e.\ that
\begin{equation}
  \frac{dN_\gamma}{dE} = \left( \frac{E}{E_f} \right)^{-\alpha}
\end{equation}
where $E_f=0.1$ is the fiducial energy at which we calculate the luminosity $P_\gamma$. Note that it is probably unrealistic to assume that the spectrum continues to be a power law to higher energies (above a few hundred GeV), instead we should expect a cut or a tilt in the spectrum. However, we will here for simplicity assume that there is no cut-off which means that we will probably overestimate the diffuse gamma ray background at high energies.

\subsection{Flux from a single source}

When taking resolved blazars into account we need the gamma ray flux a given blazar would produce. A blazar with luminosity $P_\gamma$ and spectral index $\alpha$ at redshift $z$ will give rise to the integrated gamma ray flux above energy $E_{th}$,
\begin{equation}
  \Phi(E_0>E_{th}) = \frac{P_\gamma}{2 \pi \hbar E_f} \frac{E_{th}}{\alpha-1}
  \left( \frac{E_{th}(1+z)}{E_f}\right)^{-\alpha}
  \frac{1}{(R_0 S_k(r))^2}.
  \label{eq:gammaint}
\end{equation}
This equation is valid for $E_{th} \lsim 10$ GeV since we here have neglected absorption (which is a reasonable approximation for low energies).
With appropriate unit conversions this is the flux in units of cm$^{-2}$ s$^{-1}$ that should be compared with the EGRET or GLAST point source sensitivity. For EGRET we will use the point source sensitivity $1\cdot 10^{-7}$ cm$^{-2}$ s$^{-1}$ \cite{egret-pss} and for GLAST we will use $1.6\cdot 10^{-9}$ cm$^{-2}$ s$^{-1}$ \cite{Ritz-glast}.

\subsection{Distribution of spectral indices}

\begin{figure}[t]
\centerline{\includegraphics[width=0.49\textwidth]{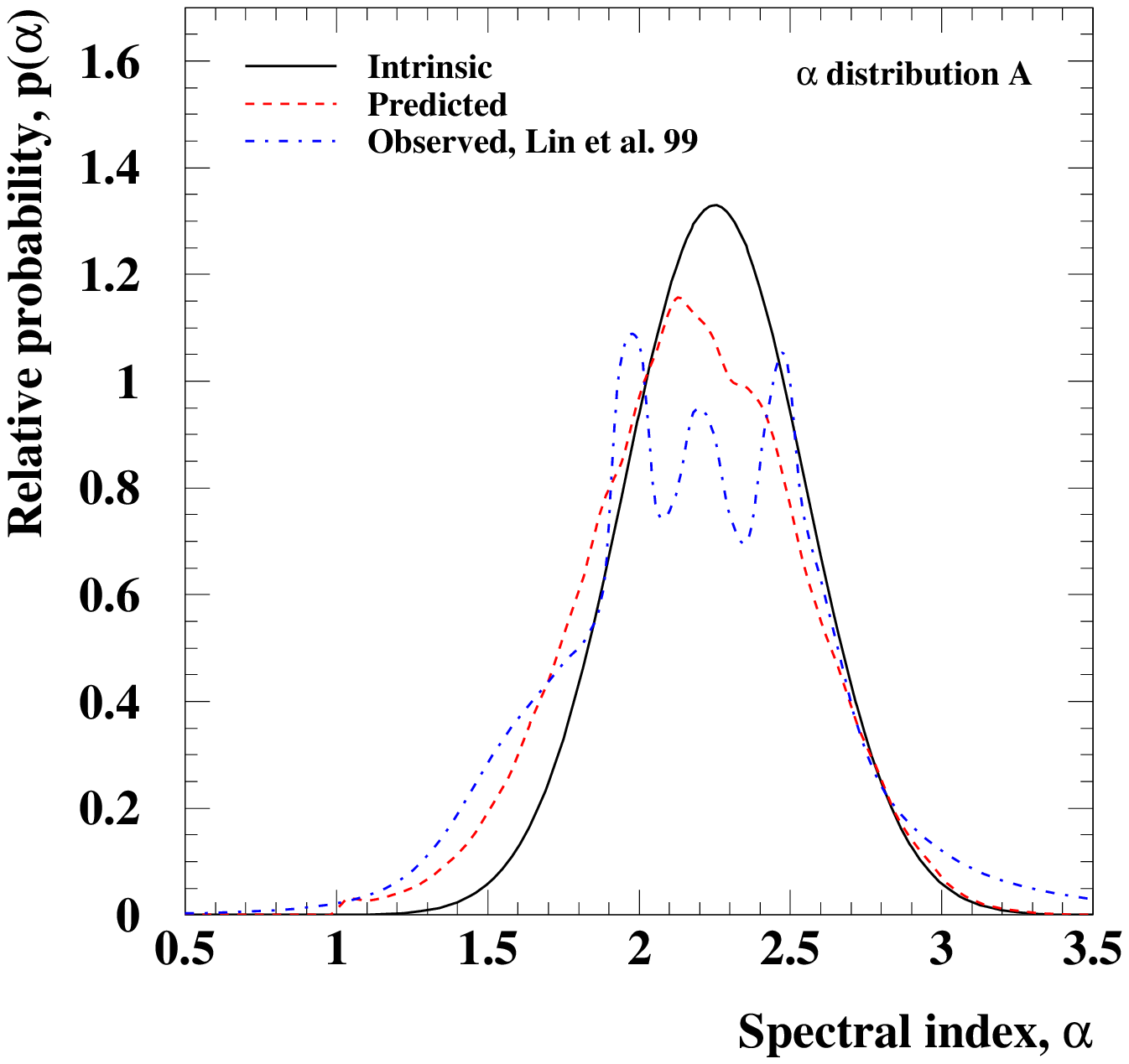}
\includegraphics[width=0.49\textwidth]{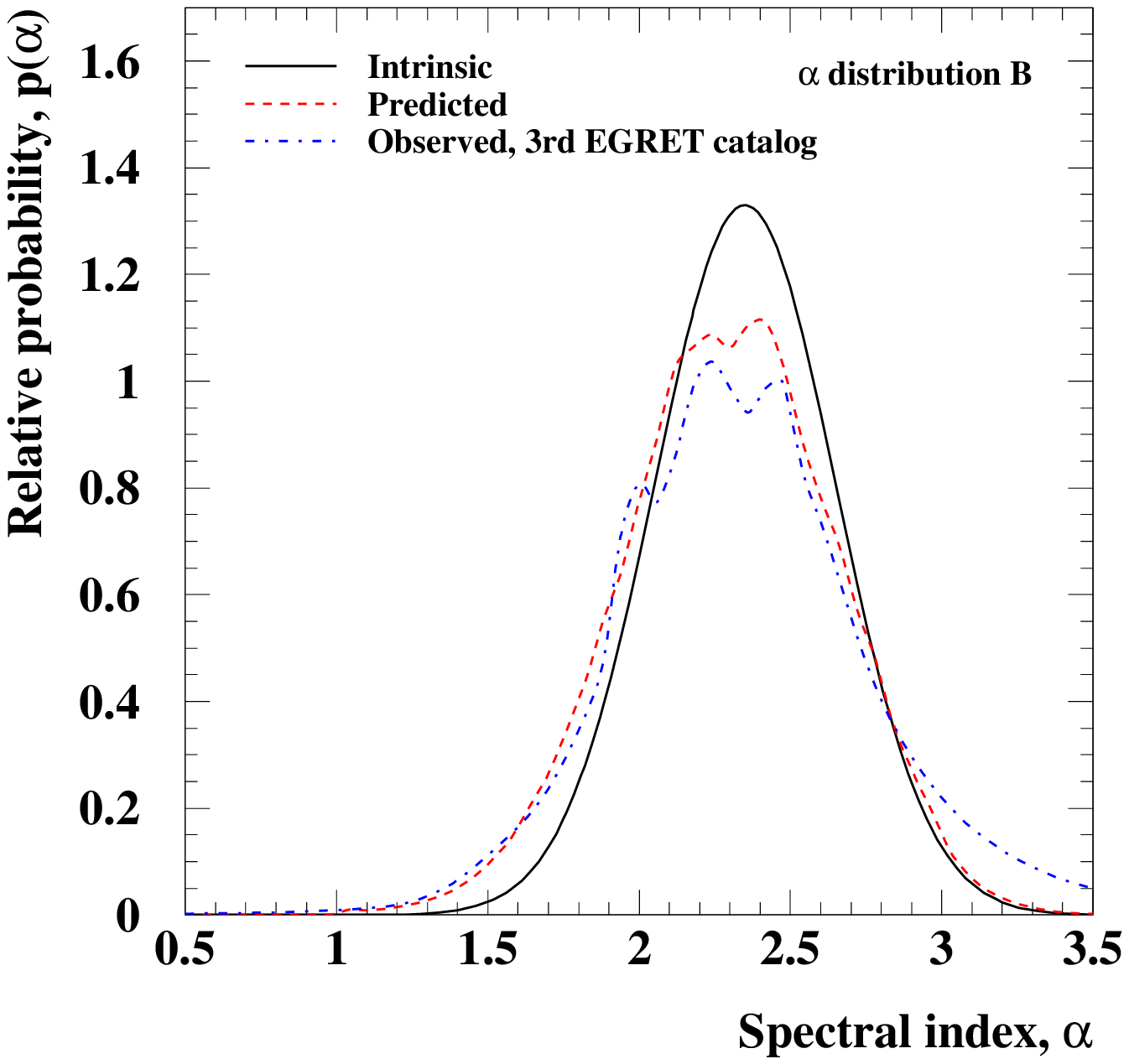}}
\caption{Different $\alpha$ distributions. a) The solid line is the intrinsic distribution A discussed in the text, the dashed is the predicted EGRET observable distribution and the dash-dotted line is the observed EGRET distribution for the Lin et al.\ sample \protect\cite{Lin99}. b) The same as in a) but for intrinsic $\alpha$ distribution B and compared with the sample in the 3rd EGRET catalog \protect\cite{Hartman99}.}
\label{fig:palpha}
\end{figure}

We have to choose a distribution function for the spectral indices, $p(\alpha)$. One option is to use the distribution of spectral indices of blazars as observed by EGRET,
\begin{equation}
  p(\alpha) = \frac{1}{N} \sum_i^N \frac{1}{\sigma_i \sqrt{2\pi}} 
   e^{-\frac{(\alpha-\alpha_i)^2}{2 \sigma_i^2}}
\end{equation}
where the sum is over the observed spectral indices $\alpha_i$ with their corresponding errors $\sigma_i$. However, this is not the best choice of distribution function since sources with low $\alpha$ are easier to detect due to their harder spectrum and we would hence introduce a selection bias. Instead we select a distribution function of the form
\begin{equation}
  p(\alpha) = \frac{1}{\sigma_{\rm int} \sqrt{2\pi}} 
   e^{-\frac{(\alpha-\alpha_{\rm int})^2}{2 \sigma_{\rm int}^2}}
\end{equation}
where we fix $\alpha_{\rm int}$ and $\sigma_{\rm int}$ such that the predicted distribution of $\alpha$ for observable blazars match the observed distribution. The predicted $\alpha$ distribution as it should be observed by EGRET is given by
\begin{eqnarray}
   p_{\rm obs}(\alpha) d\alpha & = & 
  \frac{1}{N_{\rm pred}} \frac{4 \pi c}{H_0}
  \int_{\log_{10} P_\gamma^{\rm min}}^{\log_{10} P_\gamma^{\rm max}} d\log_{10}{P_\gamma}
  \int_{z_{\rm min}}^{z_{\rm max}} dz \nonumber \\
  & & 
  \frac{1}{h(z)} \Big[
   p_{\rm int}(\alpha-\Delta\alpha_q) \rho_{\gamma,q}(P_\gamma,z)+
   p_{\rm int}(\alpha-\Delta\alpha_f) \rho_{\gamma,f}(P_\gamma,z) \Big]
   \left(R_0 S_k(r)\right)^2
   d\alpha.
  \label{eq:alobs}
\end{eqnarray}
where we only integrate over observable blazars, which is most easily done by noting that a blazar at redshift $z$, with luminosity $P_\gamma$ and spectral index $\alpha$ is observable if it would produce a flux above the EGRET point source sensitivity of $~1 \cdot 10^{-7}$ cm$^{-2}$ s$^{-1}$ integrated above 0.1 GeV\@.
Using Eq.~(\ref{eq:gammaint}) above we can for a given $P_\gamma$ and $\alpha$ calculate the maximum redshift $z'$ at which such a blazar would be observable. This would then be our upper limit for the $z$-integration, i.e.\ $z_{\rm min}=0$ and $z_{\rm max}=z'$. $N_{\rm pred}$ in Eq.~(\ref{eq:alobs}) is the total number of observable blazars and is given by
\begin{eqnarray}
   N_{\rm pred} & = & 
  \frac{4 \pi c}{H_0}
  \int_{\log_{10} P_\gamma^{\rm min}}^{\log_{10} P_\gamma^{\rm max}} d\log_{10}{P_\gamma}
  \int_{\alpha_{\rm min}}^{\alpha_{\rm max}} d\alpha
  \int_{z_{\rm min}}^{z_{\rm max}} dz \nonumber \\
  & & 
  \frac{1}{h(z)} \Big[
   p(\alpha-\Delta\alpha_q) \rho_{\gamma,q}(P_\gamma,z)+
   p(\alpha-\Delta\alpha_f) \rho_{\gamma,f}(P_\gamma,z) \Big]
   \left(R_0 S_k(r)\right)^2
  \label{eq:nobs}
\end{eqnarray}
We now have to choose a sample of observed blazars and fit $\alpha_{\rm int}$ and $\sigma_{\rm int}$ such that we can reproduce the observed distribution of $\alpha$. We have followed this procedure for two samples of blazars, the first one are 27 blazars by Lin et al.\ \cite{Lin99} and the second one are the 65 blazars with determined spectral indices in the 3rd EGRET catalog \cite{Hartman99}. Before we do this fit, we fix the spectral shifts of blazars in quiescent and flaring states as 
\begin{equation}
  \left\{ \begin{array}{lcl}
  \Delta\alpha_q & = & -0.05 \\
  \Delta\alpha_f & = & 0.20 \\
  \end{array}\right.
\end{equation}
which are the shifts determined by Stecker \& Salamon \cite{StSa96} for EGRET blazars which have been observed in both quiescent and flaring states. For the two samples we then get
\begin{equation}
  \left\{ \begin{array}{lcl}
  \alpha_{\rm int}^A & = & 2.25 \\
  \sigma_{\rm int}^A & = & 0.30 \\
  \end{array} \right. \quad \mbox{Lin et al.\ \cite{Lin99}}
  \qquad ; \qquad
  \left\{ \begin{array}{lcl}
  \alpha_{\rm int}^B & = & 2.35 \\
  \sigma_{\rm int}^B & = & 0.30 \\
  \end{array} \right. \quad \mbox{3rd EGRET catalog \cite{Hartman99}}
\end{equation}
These values are in very good agreement with the results in \cite{Sreekumar01}. We will refer to the first and second set of parameters as distribution A and B respectively. In Fig.~\ref{fig:palpha}a we compare distribution A with the predicted EGRET distribution and the observed EGRET distribution. In Fig.~\ref{fig:palpha}b we do the same thing for distribution B\@. Note that both predicted distributions fit the two observed samples rather well, but that the sample in the 3rd EGRET catalog is shifted by about 0.1 compared to the Lin et al.\ sample. This shift is of the same order as the expected systematic uncertainty in the EGRET catalog. In the following, we will use distribution A as our default since it reproduces the EGRET observed diffuse extragalactic background better than distribution B (see section \ref{sec:resolved} below). 

The predicted number of observed blazars is given by Eq.~(\ref{eq:nobs}) and for the two distributions we get $N_{\rm pred}^A=51$ and $N_{\rm pred}^B=42$, in reasonable agreement with the observed number of 66 blazars \cite{Hartman99}. Note that we do not expect perfect agreement since we only use a simple point source sensitivity, but in reality the sensitivity is much more complicated. We could easily envision that it should depend on e.g.\ the spectral index $\alpha$. Hence we are content that the agreement is as good as it is. Note that we also have the freedom to change the beaming parameter $\eta$, but we choose to keep it fixed to $\eta=1$ as given in Ref.\ \cite{StSa96}. 

\subsection{Taking resolved blazars into account}
\label{sec:resolved}

In Eq.~(\ref{eq:gammabkg}) we should only integrate over unresolved blazars. This is done in the same way as in the previous section, i.e.\ for a given luminosity $P_\gamma$ and spectral index $\alpha$ there is a given redshift $z'$ below which the blazars will be resolved and above which they will be unresolved. If we let the lower limit of the redshift integration, $z_{\rm min}$ be equal to this redshift $z'$ we will only include unresolved blazars.

\begin{figure}
\centerline{\includegraphics[width=0.49\textwidth]{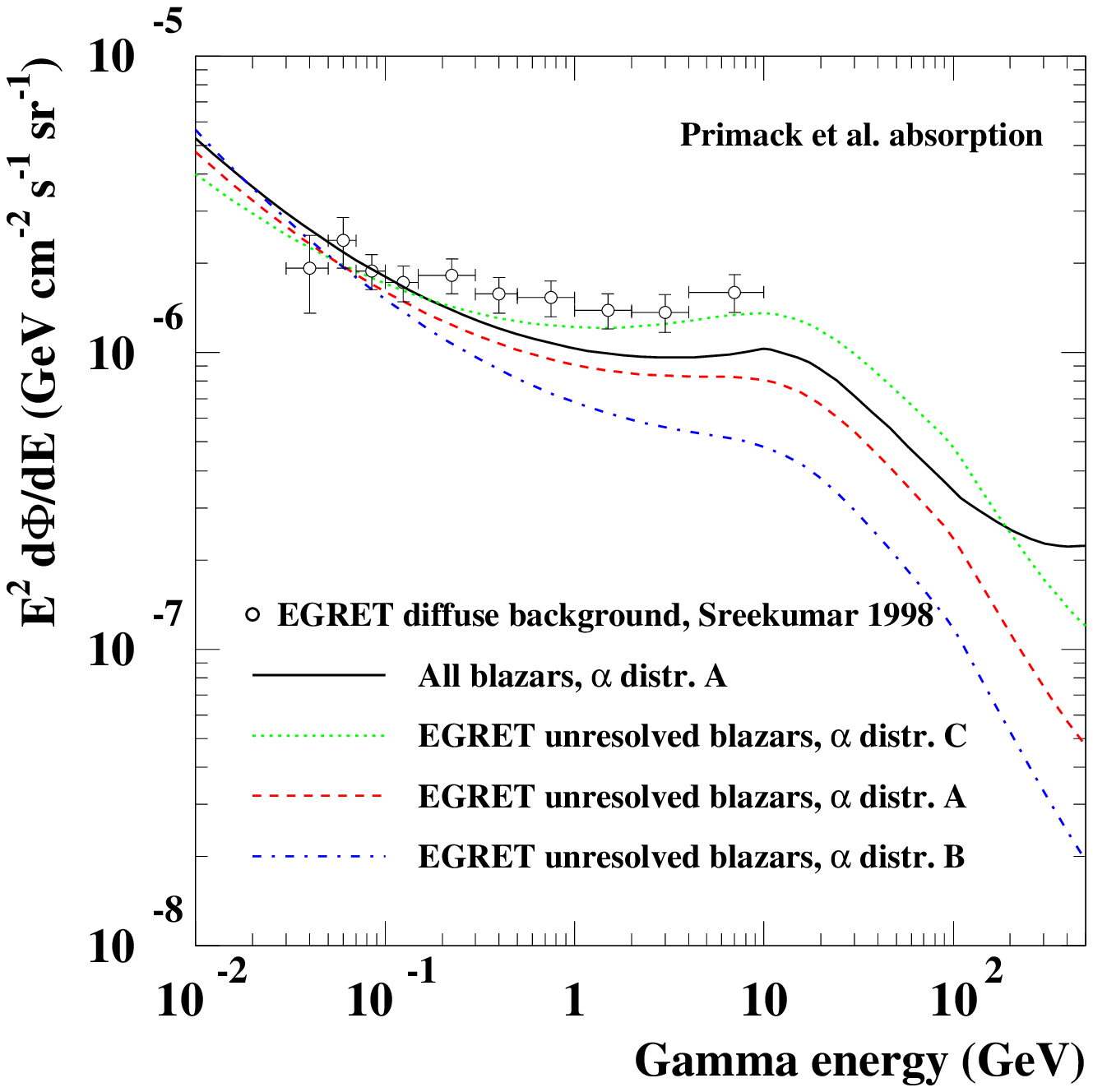}
\includegraphics[width=0.49\textwidth]{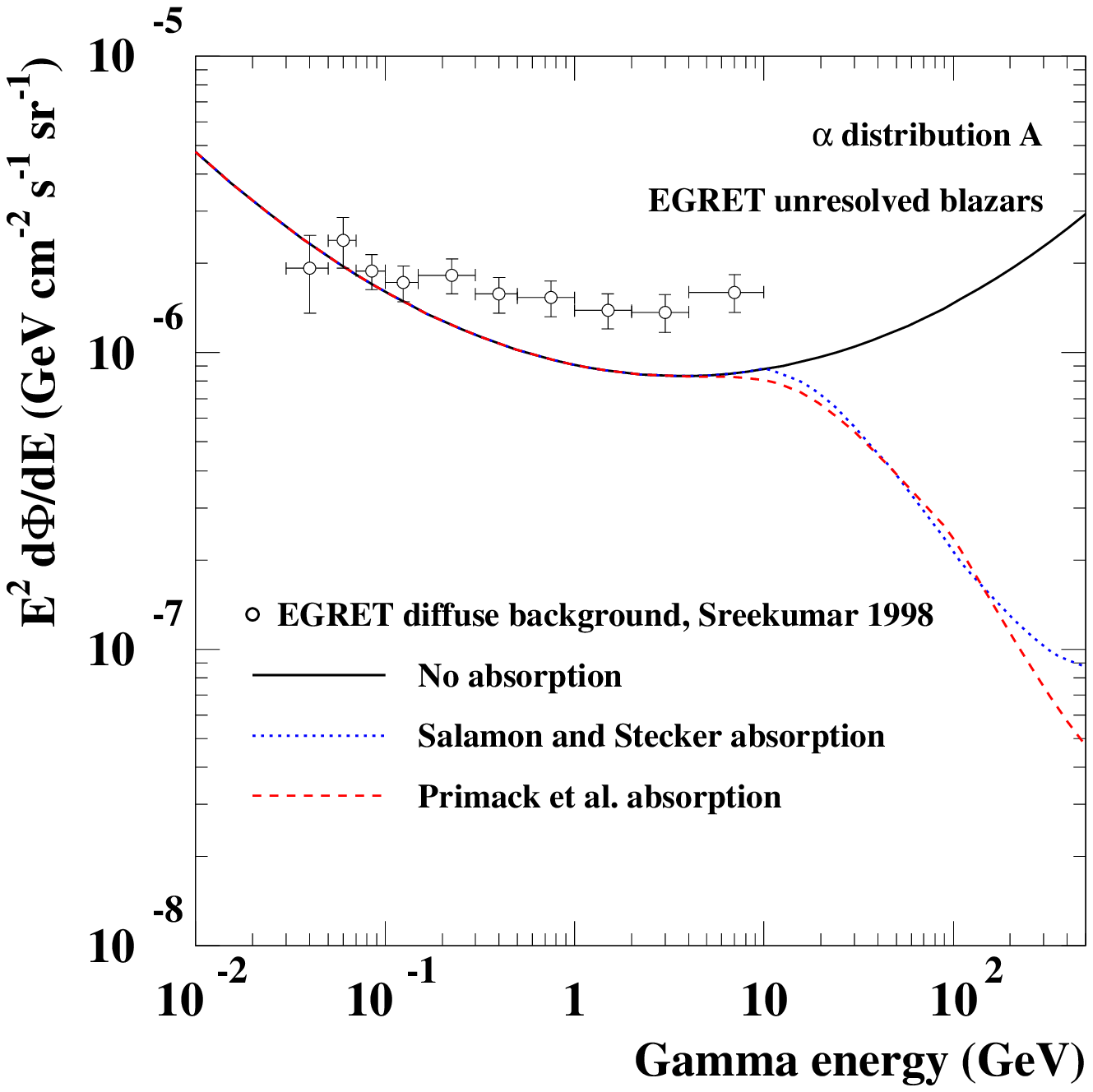}}
\caption{The predicted diffuse gamma ray flux (multiplied by $E^2$ to show features more clearly) for EGRET. a) The predicted fluxes for different $\alpha$ distributions. Distribution C is like A and B but with $\alpha_{\rm int}=2.15$. As can be seen the exact shape of the spectrum is fairly sensitive to the $\alpha$ distribution. Also shown are the EGRET measurements of the diffuse extragalactic gamma ray background. b) The predicted flux for different absorption models.}
\label{fig:e2dfde-egret}
\end{figure}

In Fig.~\ref{fig:e2dfde-egret} we show the predicted diffuse gamma ray fluxes for EGRET\@. As can be seen in a), our model reproduces the measured diffuse extragalactic background \cite{Sreekumar98} fairly well. To further illustrate the dependence on the $\alpha$ distribution, we also show results for a hypothetical $\alpha$ distribution (C) with $\alpha_{\rm int}=2.15$. For this distribution, the agreement with the EGRET measurements is excellent (the slight difference in normalization could be fixed by slightly increasing the beaming parameter $\eta$). We should note that our predictions are fairly sensitive to the exact low-$\alpha$ behavior of $p(\alpha)$. The higher up in energy we go, the more we sample the low-$\alpha$ region. In b) we show the effects of the different absorption models. It is clear that as soon as we go above 10--100 GeV, absorption effects are very important. Also keep in mind that we have not included a cut-off in the intrinsic gamma ray spectrum which would further reduce the fluxes at high energy.

\begin{figure}
\centerline{\includegraphics[width=0.49\textwidth]{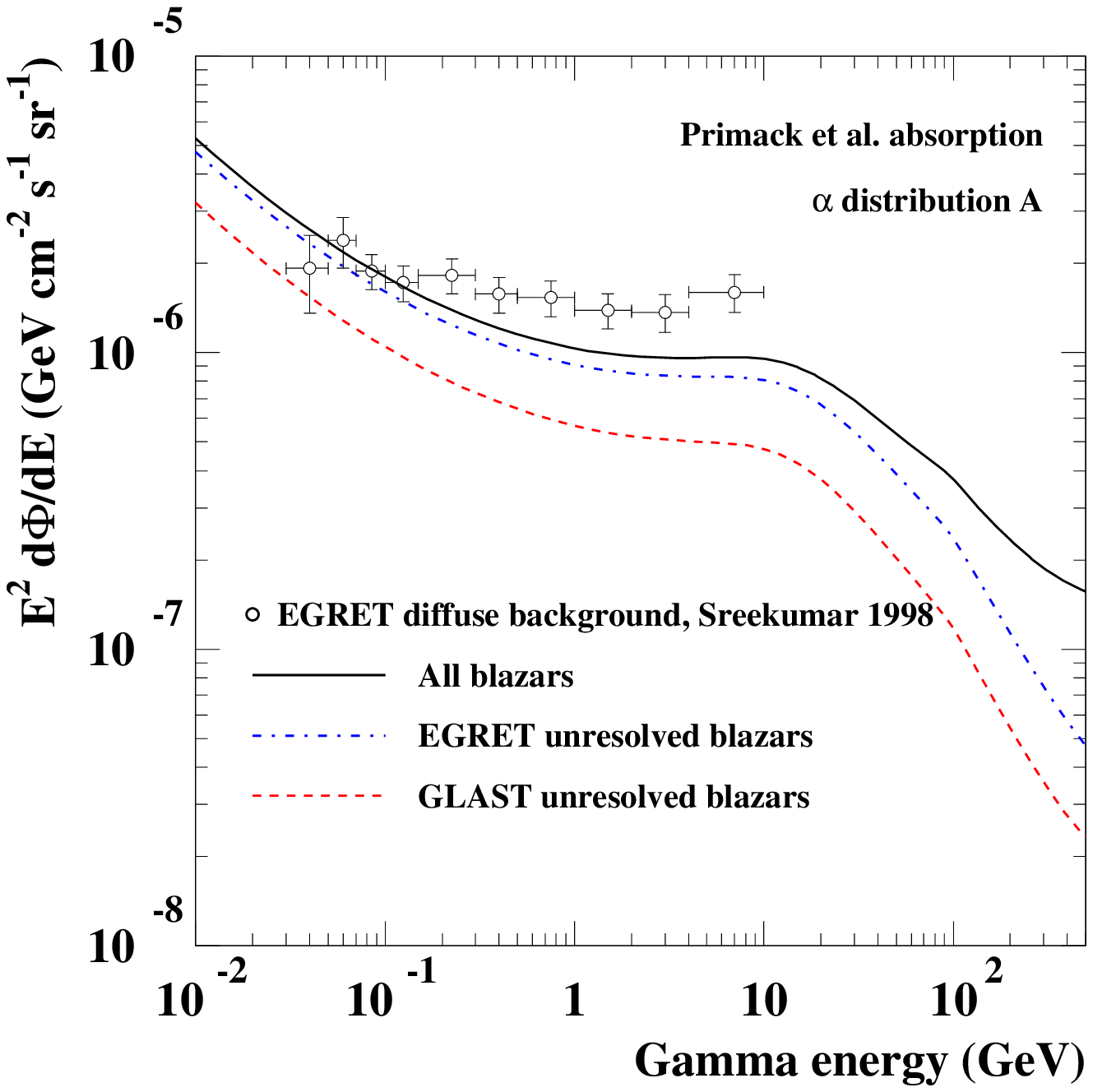}
\includegraphics[width=0.49\textwidth]{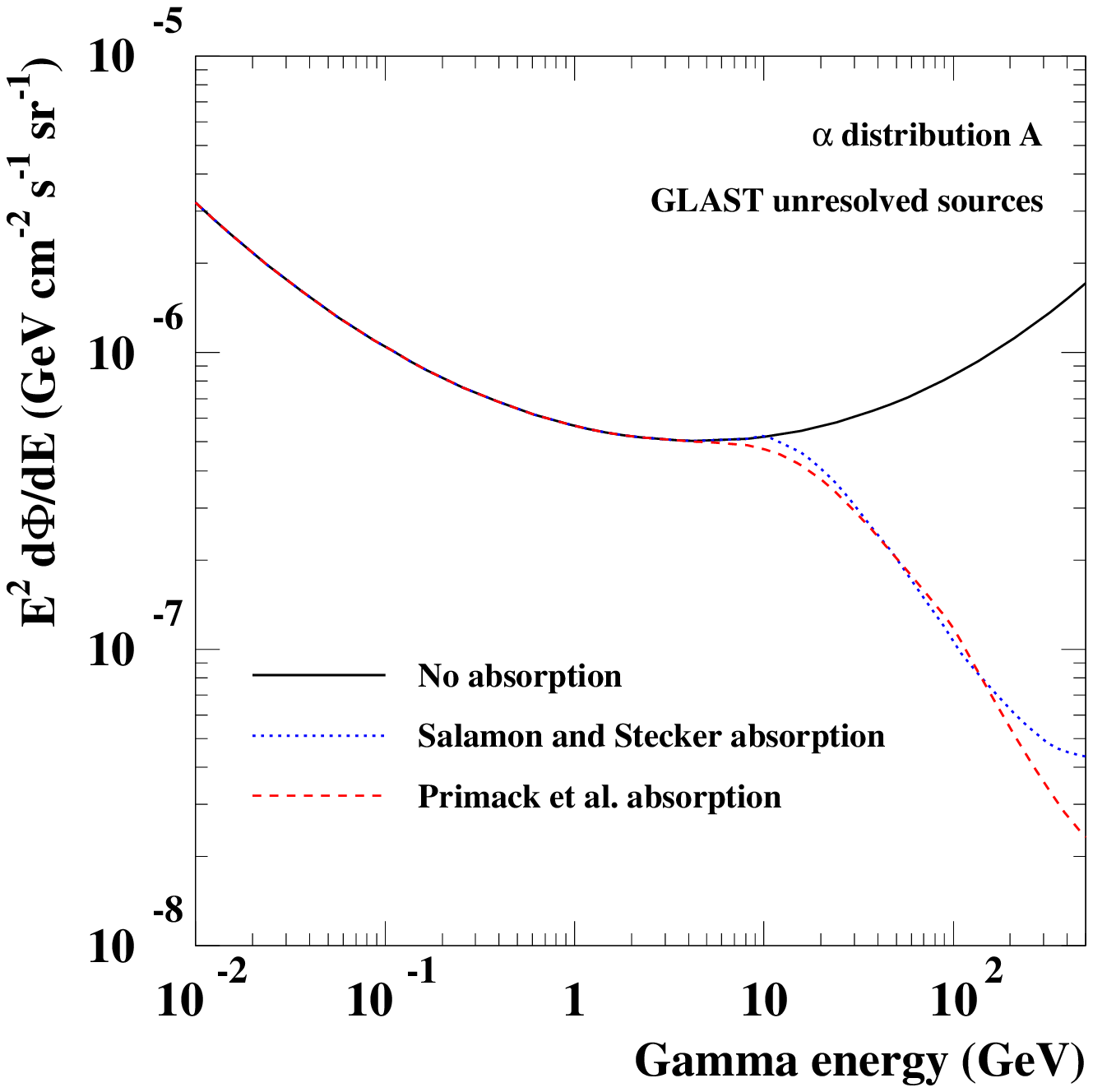}}
\caption{The predicted diffuse gamma ray flux (multiplied by $E^2$ to show features more clearly) for GLAST. a) The predicted fluxes compared to the EGRET measurements. b) The predicted fluxes for different absorption models.}
\label{fig:e2dfde-glast}
\end{figure}

In Fig.~\ref{fig:e2dfde-glast}a we show the effect of different point source sensitivities. We see that compared to EGRET, the superior point source sensitivity of GLAST will reduce the diffuse gamma ray background with roughly a factor of two. Note however, that the angular resolution of GLAST will make the point source sensitivity worse at lower energies (or rather, larger spectral indices $\alpha$), an effect we have not included here. We expect that this effect would make the predicted background for GLAST slightly higher at low energies than shown in the figure. In Fig.~\ref{fig:e2dfde-glast}b we show the effect of different absorption models for the predicted GLAST gamma ray background.

\subsection{Uncertainties}

In this section we have produced a derivation of the expected diffuse gamma ray background assuming that it is due to unresolved blazars. There are many uncertainties involved. First of all, it is not known whether blazars are the only sources relevant to compute the background. The energy spectrum of the blazars is also not very well known, i.e.\ there could be a cut-off at high energies (and even if the spectrum is a power law, the distribution of spectral indices is uncertain). Even the luminosity function is rather uncertain and the assumption of the relation between the gamma and radio luminosity functions cannot be tested until the sample of blazars measured in both gamma and radio increases. The parameters of the model we discussed are also quite uncertain, and, as already mentioned, gamma ray absorption introduces further uncertainties, especially at high energies. In spite of all these uncertainties, the agreement we find between our prediction and EGRET data is quite good, and gives some credibility to our estimate of the background for GLAST at higher energies. We have chosen as our default model the $\alpha$ distribution A, which reproduces both the measured $\alpha$ distribution and the EGRET energy spectrum satisfactory, and the absorption model of Primack et al. Keep in mind though that, above $\sim100$ GeV, the uncertainties can be as large as a factor of a few.

\section{Applications to supersymmetric dark matter} \label{sec:applications}

\subsection{A few examples in a specific particle physics setup}

So far, we have kept the discussion as general as possible, without specifying
the exact identity of the WIMP making up the dark matter. To gauge the
possibility of detecting a gamma-ray signal in a realistic scenario, we now 
investigate one of the prime candidates for dark matter: the lightest 
supersymmetric particle (LSP) in the MSSM - the minimal supersymmetric 
extension of the Standard Model of particle physics.   
If R-parity is conserved, the LSP is stable; furthermore,
its coupling with lighter standard model particles ensures that a population
of such particles is present in the early Universe, with its density set by
thermal equilibrium. The freeze out from equilibrium is roughly set by the LSP
thermally averaged annihilation cross section; as sketched in 
Eq.~(\ref{eq:scal}),
a weak interaction strength coupling ensures that WIMPs have a thermal
relic abundance of the order of the critical density: this is naturally 
the case if the LSP has zero electric and color charges.

We thus take as our template WIMP dark matter candidate the lightest 
neutralino, $\tilde{\chi}^0_1$, in the MSSM (see \cite{lbreview} for a 
recent review). 
$\tilde{\chi}^0_1$ is defined as the lightest mass eigenstate obtained 
from the superposition of four spin-1/2 fields, the Bino and Wino gauge 
fields, $\tilde{B}$ and $\tilde{W}^3$, and two neutral CP-even Higgsinos,
$\tilde{H}^0_1$ and $\tilde{H}^0_2$:
\begin{equation}
  \tilde{\chi}^0_1 = 
  N_{11} \tilde{B} + N_{12} \tilde{W}^3 + 
  N_{13} \tilde{H}^0_1 + N_{14} \tilde{H}^0_2\;.
\end{equation}
There are large regions in the MSSM parameter space where Bino-like LSPs
or neutralinos with relevant Higgsino components have a thermal relic 
abundance of the right order to account for the dark matter, see, 
e.g.,~\cite{coann}. We have used the DarkSUSY package~\cite{ds} to 
scan extensively the parameter space and generate a large archive
of such models. We select those models that do not violate present 
accelerator and astrophysical limits and study what is the typical
gamma-ray yield, both for the continuum and monochromatic spectra
(in the MSSM there are two line signals allowed: $\gamma\gamma$  
and $Z\gamma$). With DarkSUSY \cite{ds} we calculate the relic density by 
numerically solving the Boltzmann equation properly taking resonances, 
thresholds and coannihilations (between the lightest neutralinos and other 
neutralinos and charginos) into account \cite{coann}.

\begin{figure}[t]
\centerline{
\includegraphics[width=0.70\textwidth]{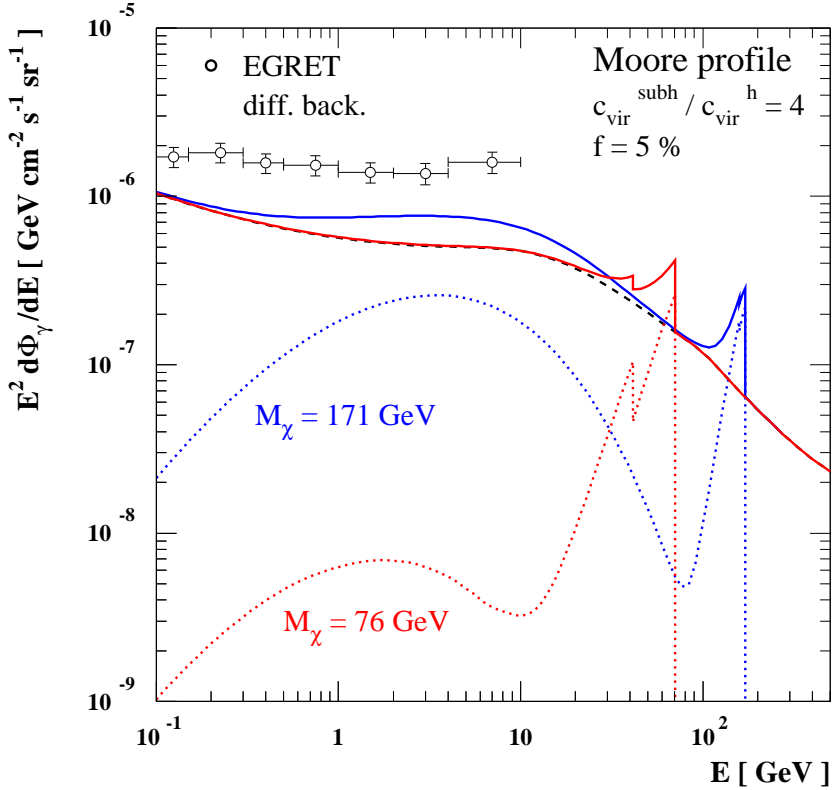}} 
\caption{Extragalactic gamma-ray flux (multiplied by $E^2$) for two sample 
thermal relic neutralinos in the MSSM (dotted curves), summed 
to the blazar background expected for GLAST (dashed curve). 
Normalizations for the signals are computed assuming halos are modelled by 
the Moore profile, with 5\% of their mass in substructures with 
concentration parameters 4 times larger than $c_{vir}$ as estimated
with the Bullock et al.\ toy model.} 
\label{fig:mssm}
\end{figure}

\begin{figure}[t]
\centerline{
\includegraphics[width=0.70\textwidth]{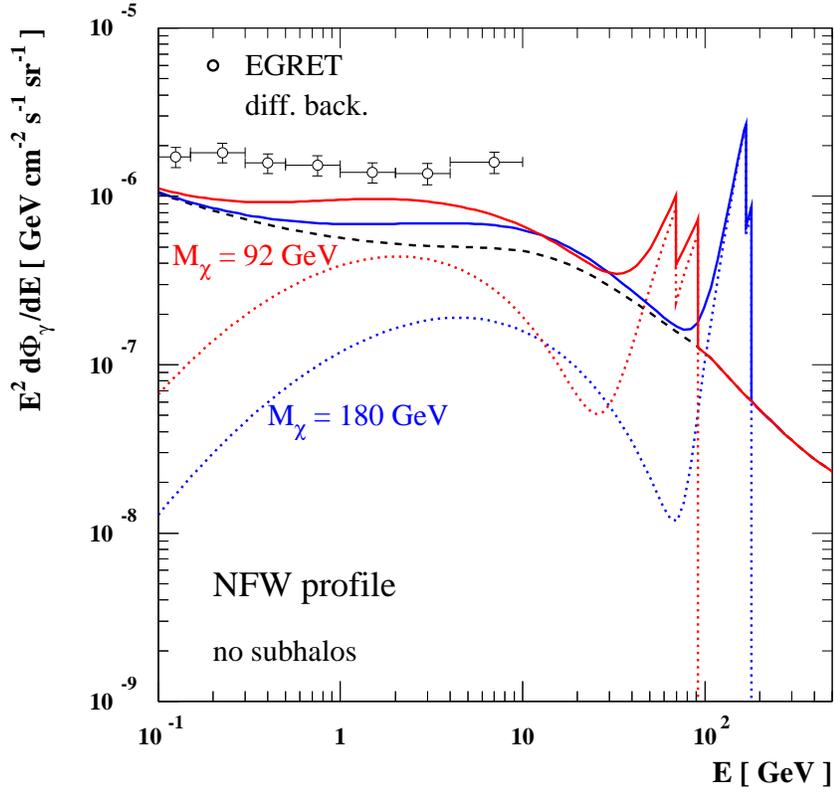}} 
\caption{Extragalactic gamma-ray flux (multiplied by $E^2$) for two sample 
non-thermal dark matter candidates arising in the AMBS scenario (dotted
lines) compared to the expected background (dashed curve). Annihilations 
cross sections are in these cases larger than for the models displayed 
in Fig.~\protect{\ref{fig:mssm}}, however a different normalization for the 
fluxes is implemented here: we consider the case for halos modelled by 
the NFW profile, no substructures and concentration parameters inferred
from the Bullock et al.\ toy model.}
\label{fig:amsb}
\end{figure}

We focus first on neutralinos with a relic abundance $\Omega_\chi h^2$ in 
the interval 0.1 to 0.2, corresponding to our preferred cosmology
$\Omega_M \sim 0.3$ and $h \sim 0.7$. There are models with
$\sigma v_{2\gamma} \gtrsim 10^{-29} \rm{cm}^3 \rm{s}^{-1}$ over the 
whole mass range from 50 GeV up to a few TeV. We consider two sample cases
and plot the corresponding extragalactic $\gamma$-ray spectra
in Fig.~\ref{fig:mssm} (dotted lines). The first
model has $M_\chi$~=~76~GeV, relatively low total annihilation cross 
section $\sigma v = 6.1\cdot 10^{-28} \rm{cm}^3 \rm{s}^{-1}$ but large 
branching ratios into photon states, $b_{2 \gamma} = 6.1\cdot 10^{-2}$ and 
$b_{Z \gamma} = 5.2\cdot 10^{-2}$. The other has $M_\chi$~=~171~GeV, larger
annihilation rate $\sigma v = 4.5\cdot 10^{-26} \rm{cm}^3 \rm{s}^{-1}$
but $b_{2 \gamma} = 5.2\cdot 10^{-4}$ and negligible branching ratio into 
the $Z\gamma$ final state.
The normalization of the flux is set by assuming that dark matter
structures are described by the Moore profile, with concentration 
parameters as computed with the Bullock toy model, and assuming the 
presence of a moderate amount of substructure, $f = 5\%$, with a factor 
of 4 enhancement in $c_{vir}$. Under these assumptions, the neutralino 
induced $\gamma$-ray flux is at the level of the diffuse background from 
unresolved blazars ($\alpha$ distribution A) expected in GLAST, with the 
peak from the monochromatic emission significantly above it (dashed curves 
refer to the background only, solid curves to the sum of signal plus 
background). 

The condition $\Omega_\chi h^2 > 0.1$ sets an upper bound on to the total 
annihilation cross section and hence, indirectly, an upper bound on the
strength of the monochromatic channels; such states however are not
the dominant modes and therefore a lower bound does not follow from imposing
$\Omega_\chi h^2 < 0.2$: there are cases where the $\tilde{\chi}^0_1$
is compatible with being a good dark matter candidate, but the monochromatic
flux is negligible. An opposite conclusion holds for the continuum 
components: there are cases in which the gamma-ray yield can be slightly 
larger than the one for the $M_\chi$~=~171~GeV model, but very small regions
in parameter space where the yield is significantly smaller than for
the model with $M_\chi$~=~76~GeV we show.  

If we remove the constraint on $\Omega_\chi$ the picture can change 
drastically. In particular, there are several schemes in which the 
LSP relic abundance today is not set by its thermal relic density.
One example is the case for Wino or Higgsino-like neutralinos in the 
version of the MSSM with anomaly mediation for supersymmetry breaking 
(AMSB). This scheme induces a mapping into regions in the MSSM parameter 
space in which the thermal relic abundance is negligible; however, an 
additional ``non-thermal'' relic source is present due to decays into 
neutralinos of gravitinos or moduli fields, fields that parameterize a 
flat direction of the theory and that dominate the energy density in 
the early Universe~\cite{ggw,mr}. One can show that, in this context, the 
total annihilation rate, as well as cross sections in the $2\gamma$ and 
$Z\gamma$ final states, are forced to be very large~\cite{anom}. 
Two examples are shown in Fig.~\ref{fig:amsb}:
one model has $M_\chi$~=~92~GeV, 
$\sigma v = 2.5\cdot 10^{-24} \rm{cm}^3 \rm{s}^{-1}$, 
$b_{2 \gamma} = 1.2\cdot 10^{-3}$ and $b_{Z \gamma} = 2.2\cdot 10^{-3}$; 
the second $M_\chi$~=~180~GeV,
$\sigma v = 2.2\cdot 10^{-24} \rm{cm}^3 \rm{s}^{-1}$, 
$b_{2 \gamma} = 1.8\cdot 10^{-3}$ and $b_{Z \gamma} = 5.1\cdot 10^{-3}$.
The normalization of the two extragalactic $\gamma$-ray fluxes is set 
assuming NFW halo profiles with no substructure and concentration 
parameters as computed with the Bullock toy model. Had we chosen the
Moore profile rather than NFW, the predicted fluxes would be one order
of magnitude larger, hardly compatible with the extragalactic flux as 
inferred from EGRET data. Note that a flux at roughly the same level
is expected implementing the Burkert profile, hence the detectability
of this signal is not linked to having a singular halo profile describing
dark matter halos.

\begin{figure}[t]
\centerline{
\includegraphics[width=0.70\textwidth]{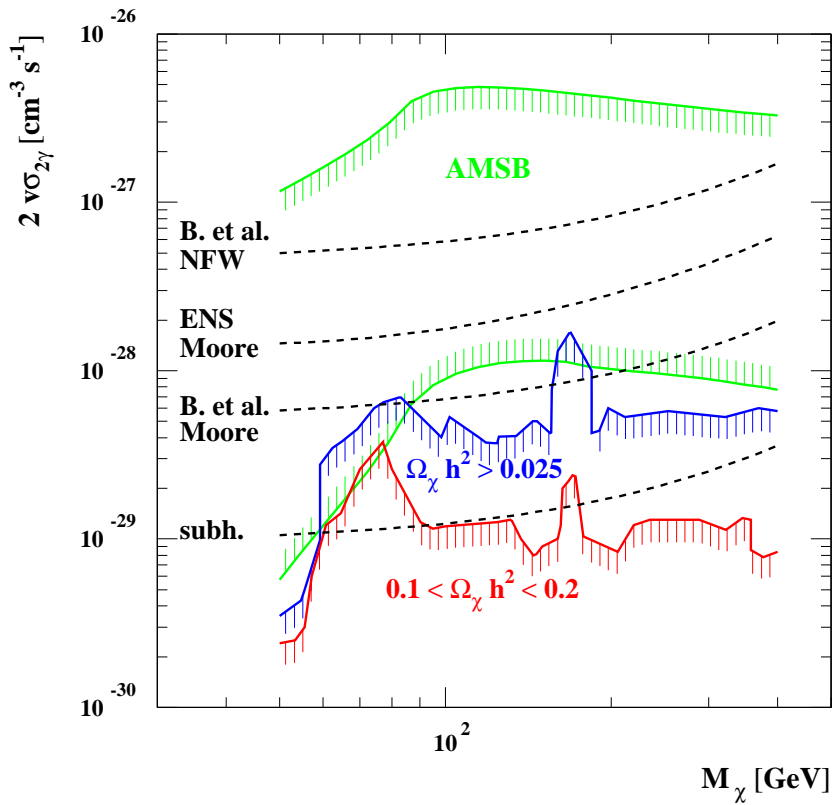}
}
\caption{Approximate $3 \sigma$ sensitivity curves for the GLAST telescope to
search for a component in the extragalactic gamma-ray flux induced by WIMP 
annihilations into monochromatic photons. The sensitivity curves are
plotted in the plane WIMP mass (coinciding with the energy peak in the 
induced flux) versus twice the annihilation rate into two photons, and
for four configurations to estimate the normalization of the flux 
(the highest and lowest dashed curves correspond, respectively, to the 
choice in Fig.~\protect{\ref{fig:amsb}} and Fig.~\protect{\ref{fig:mssm}}). 
Also shown is the range of the predictions of $v\sigma_{2\gamma}$ for
neutralinos in the AMSB scenario, and the upper limit to it in case
of thermal relic neutralinos in the MSSM, assuming their relic abundance
is either in the cosmologically preferred mass range 
$0.1 < \Omega_\chi h^2 < 0.2$, or in the less restrictive range often 
considered $0.025 < \Omega_\chi h^2 < 0.2$.} 
\label{fig:fig10}
\end{figure}

\subsection{Sensitivity in upcoming measurements}

It is not straightforward to estimate the smallest
WIMP induced component GLAST will be able to disentangle from the 
background. A firm statement about the possibility to single out the 
yield with continuum energy spectrum will be possible only when higher
precision measurements will allow to characterize better the level and
the spectral features of the background. The monochromatic component
has a much better signature and might be unambiguously identified.
We make an attempt to make a rough 
estimate of the sensitivity curves for a GLAST-like instrument under 
a few schematic assumptions. 

We assume the instrument has a peak effective area of 8000~cm$^2$
at energies above 10 GeV and an average energy resolution of 
15\%~\cite{Ritz-glast}. We take an exposure of 4~years, mapping the
whole sky except for the regions already excluded in the EGRET 
analysis~\cite{Sreekumar98}, i.e.\ the galactic plane $|b|<10^{\circ}$,
and the bulge $|l|<40^{\circ}$ and $|b|<30^{\circ}$, with an average
effective area which is about $20\%$ of the peak area. We set up
a $\chi^2$ procedure to check if we can discriminate the spectrum of a 
line signal superimposed on the background from the spectrum of 
the background only. The analysis is performed assuming a given 
normalization for the WIMP flux and keeping as free parameters the value of
the WIMP mass and annihilation cross section $\sigma v_{2\gamma}$. 
For each parameter choice, we sum to this flux the diffuse background from 
unresolved blazars ($\alpha$ distribution A) with the normalization
computed in the previous Section and already shown in 
Figs.~\ref{fig:mssm}--\ref{fig:amsb}. We then perform
a binning of the spectrum above 10~GeV, optimizing the bin width as a 
function of the number of events in each bin and checking that we have 
at least 10 events per bin. Naively, the statistical error in each bin
would be the square root of the number of events in the bin; at second 
thought though, the extragalactic background component will be obtained after 
subtracting point sources and the diffuse galactic emission, with a non 
trivial propagation of errors we cannot easily retrace here. We make a 
simplifying assumption at this stage, expecting just a rough 
estimate of the true sensitivity curves.
Suppose the main component one has to fight against is due to diffuse 
galactic $\gamma$-rays; such a component can be removed after assuming
a model for diffuse emission and should be, on average, about an order of 
magnitude larger than the extragalactic component~\cite{Sreekumar98}. 
We mimic this subtraction by assuming that the error in each energy
bin is the square root of the number of events in the bin multiplied
by 10. We then use the $\chi^2$ criterion to discriminate whether or not
the obtained distribution of events can be fitted at 3~$\sigma$ with a 
background component only with fixed spectral shape but arbitrary 
normalization. 

The corresponding sensitivity curves are shown in Fig.~\ref{fig:fig10}
in the plane neutralino mass - twice the $2\gamma$ annihilation rate.
Each curve corresponds to a different normalization for the extragalactic 
flux: from the bottom up, case for Moore profile halos with 
substructure introduced already in Fig.~\ref{fig:mssm}, 
the case for Moore profile halos with no substructure and concentration 
parameters as computed with the Bullock et al.\ toy-model or with the 
ENS model,
and, finally, the case for NFW halos with no substructure and $c_{vir}$
as in the Bullock et al.\ model. Also shown in the picture are the span 
in the predictions for $\sigma v_{2\gamma}$ in the AMSB scenario~\cite{anom}
(green lines marks the upper and lower limits for a given mass),
and approximate upper limits in the case of MSSM thermal relic neutralinos 
with relic abundance in the preferred range $0.1 < \Omega_\chi h^2 < 0.2$,
or in the less restrictive range often considered 
$0.025 < \Omega_\chi h^2 < 0.2$, as deduced from our sampling of the 
parameter space. As can be seen, depending on the configuration
one considers, there is a fair chance that the monochromatic 
$\gamma$-ray flux will be disentangled from GLAST data.
The four models we have considered in Figs.~\ref{fig:mssm} and \ref{fig:amsb}
lie all above the corresponding sensitivity curves.

The same sensitivity curve can be applied to the case of the line signals 
generated in the $Z\gamma$ channel by replacing $M_{\chi}$ on the
horizontal axis with the energy of the peak 
$E = M_{\chi} (1-M_Z^2/4\,M_\chi^2)$ and assuming the quantity on the 
vertical axis is $\sigma v_{Z\gamma} \cdot 4 E^2/(E+\sqrt{E^2+M_Z^2})^2$.

\subsection{Comparison with other signals}

It is not straightforward to compare the dark matter signal we have 
presented here with other indirect signals which were proposed soon 
after the idea of WIMP dark matter was raised, two decades ago.
Most analyses have been devoted to the study of the detectability of
gamma-rays produced in the halo of our own Galaxy or of antimatter
generated by WIMP pair annihilations taking place in our local 
environment (say within a few kpc, the exact number depends on the model for
propagation of charged cosmic rays), refining the original 
proposals, see~\cite{lbreview} for a detailed
reference list. 

As already mentioned, the prospects for detecting gamma-rays produced
in the Milky Way are much more tightened to assumptions on
the distribution of dark matter WIMPs in the galactic halo. The 
monochromatic flux generated by the sample MSSM models displayed in 
Fig.~\ref{fig:mssm} and Fig.~\ref{fig:amsb} in the Galactic center
region is within the sensitivity of GLAST or the upcoming 
generation of ground based air Cherenkov telescopes (see, e.g., the analysis
in Ref.~\cite{anom}) if indeed the dark matter density profile is singular, 
respectively, as in the Moore et al.\ or the NFW halo all the way to the 
central black hole (or maybe even steeper than that, see the possible 
enhancement induced by the black hole formation discussed in~\cite{GS}, 
but take into account also the opposite conclusions drawn in, e.g., 
Refs.~\cite{spike,milo}). As suggested also by Fig.~\ref{fig:fig4}, even 
a slight depletion in the central density can change drastically this 
conclusion. 

We checked also that the four sample MSSM models we introduced, with the 
halo profiles considered in the corresponding figures, do not generate a 
continuum spectrum component which exceeds the flux measured
by the EGRET telescope~\cite{MH}. A comparison with the 
Galactic flux at high latitudes in a configuration with clumps in the halo 
is much more uncertain. The flux is dominated by eventual nearby clumps, 
depending critically on the actual physical realization which happens to 
correspond to the Milky Way: we recall, on the other hand, that the signal 
we propose is obtained as the sum of many unresolved sources, i.e.\ 
we are automatically making an average over a set of possible
configurations. The chance for GLAST to resolve single clumps, for
the sample configurations with clumps considered in Fig.~\ref{fig:mssm}, 
can be read out of Fig.~\ref{fig:point}. The models with neutralino masses
$M_{\chi} =$~76, 171~GeV, have, respectively, $N_{\gamma}^{100} =$~27.4 and 
39.9; hence dotted lines corresponding to the Moore profile in 
Fig.~\ref{fig:point} should be rescaled along the vertical axis by, 
respectively, a factor of 0.096 and 0.45.

Limits from charged cosmic ray data are also model dependent, as again
the dark matter signal is dominated by local sources; dark matter candidates
may be excluded in some configurations, but allowed in others. Notice also 
that, especially if one focusses on the monochromatic gamma-ray component, 
such a signal is very weakly correlated to the production rate of e.g.\ 
antiprotons and positrons, see, e.g.,~\cite{clumpy}.

\section{Conclusions}\label{sec:conclusions}

We have studied predictions and the observability of the diffuse 
gamma ray signal from WIMP pair annihilations in external halos. 
We have found configurations that imply signals at a detectable level 
for GLAST, the upcoming gamma-ray space telescope, both for non-thermal 
dark matter neutralinos, such as in the anomaly-mediated supersymmetry 
breaking model, and for thermal relic neutralinos in the MSSM. 
The key ingredient to show that detectable fluxes may arise, which
was neglected in early estimates, is the picture, inspired by the current 
theory for structure formation and by N-body simulation results, 
that dark matter clusters hierarchically in larger and larger halos, with 
light structures more concentrated than more massive ones.
For dark matter candidates in the AMSB scenario our conclusion holds 
independently of further assumptions on the dark matter distribution 
inside halos. 
Pair annihilation cross sections for thermal relic WIMPs are generally 
smaller; this however can be compensated by the enhancement in the flux one 
finds if, as suggested by results of simulations, we assume that dark matter 
profiles are singular and contain small dense substructures.

If the branching ratio for WIMP annihilation into monochromatic gamma rays 
is significant (about a few times $10^{-4}$ or larger), the induced 
extragalactic flux shows a very distinctive feature, the asymmetric distortion
of the line due to the cosmological redshift and its sudden drop at
the value of the WIMP mass. The component with continuum energy spectrum
can be at the level of background components but has less distinctive 
features: the flux is characterized by the ``$\pi^0$ bump'', rather than
by a spectral index, with the peak shifted to energies lower than 
$M_{\pi}/2$ and the width set by the WIMP mass. Once a better
measurement of the background will be available, it will be possible to
address the question of whether or not this signal can be disentangled
from other eventual components.

We have discussed in detail how our predictions depend on assumptions
on the Cosmological model and the structure formation picture applied.
Unless one introduces drastic changes, such as a large warm dark matter 
component, the cosmological parameters in the CDM setup do not play a major 
role; results are mainly sensitive to $\sigma_8$ with about a factor of two
uncertainty. Larger indeterminations, of the order of a factor of a few, 
are introduced when estimating the scaling of the concentration 
parameter with halo mass, as extrapolations with toy models 
out of the mass range of N-body simulation results are needed. 
The functional form of the dark matter profile in single halos introduces 
a factor of 10 uncertainty, going from the case of a $1/r^{1.5}$ cusp in 
the Moore profile to the case of non-singular profiles; that uncertainty
is much smaller than, e.g., the one induced on the estimate of the flux
from the center of our own Galaxy.
The presence of substructures inside halos may provide a factor of 
a few enhancement in the flux, but this effect is more difficult to 
address: we have presented a simple and rather generic setup, which will 
be possible to refine when further information on halos will be provided 
by higher resolution numerical simulations.

Issues related to the estimate of the background are very important as well.
We have presented here a novel estimate of the expected background from
unresolved blazars in GLAST, exploiting recent data and discussing
critically the uncertainties involved, including the role played by
gamma absorption.

Concluding, the present analysis has been devoted to examining in detail an
idea that three of the authors have recently presented in a short 
letter~\cite{beu}. This work provides further support for such proposal,
with exciting perspectives for upcoming measurements.

P.U. was supported by the RTN project under 
grant HPRN-CT-2000-00152. 
J.E. and L.B. wish to thank the Swedish Research Council for support.

\end{document}